\documentclass[a4paper,11pt]{article}
\usepackage{latexsym}
\usepackage{amssymb}
\usepackage{amsmath}
\usepackage{cite}
\usepackage{color}

\usepackage{microtype}

\numberwithin{equation}{section}
\allowdisplaybreaks

\makeatletter \@addtoreset{equation}{section} \makeatother

\def\ii{{\rm i}}

\usepackage{ytableau}

\ytableausetup{centertableaux, mathmode, smalltableaux}

\definecolor{blue-violet}{rgb}{0.54, 0.17, 0.89}
\definecolor{PineGreen}{cmyk}{0.92, 0, 0.59, 0.25}
\definecolor{YellowOrange}{cmyk}{0, 0.42, 1, 0}

\newcommand{\be}{\begin{equation}}
\newcommand{\ee}{\end{equation}}
\newcommand{\beq} {\begin{equation}}
\newcommand{\eeq} {\end{equation}}
\newcommand{\ba}{\begin{eqnarray}}
\newcommand{\ea}{\end{eqnarray}}

\usepackage{graphicx}

\begin{document}
\numberwithin{equation}{section}

\begin{center}
{\bf\LARGE Metric-Affine Myrzakulov Gravity Theories} \\
\vskip 2 cm
{\bf \large Nurgissa Myrzakulov$^{\ast,\dag}$, Ratbay Myrzakulov$^{\ast,\dag}$, Lucrezia Ravera$^{\star,\ddag}$}
\vskip 8mm
 \end{center}
\noindent {\small $^{\ast}$ \it Eurasian International Centre for Theoretical Physics, Nur-Sultan 010009, Kazakhstan. \\
$^{\dag}$ \it Eurasian National University, Nur-Sultan 010008, Kazakhstan. \\
$^{\star}$ \it DISAT, Politecnico di Torino, Corso Duca degli Abruzzi 24, 10129 Torino, Italy. \\
$^{\ddag}$ \it INFN, Sezione di Torino, Via P. Giuria 1, 10125 Torino, Italy.
}

\begin{center}
\today
\end{center}

\vskip 2 cm
\begin{center}
{\small {\bf Abstract}}
\end{center}

In this paper we review the Myrzakulov Gravity models (MG-N, with $\mathrm{N = I, II, \ldots, VIII}$) and derive their respective metric-affine generalizations (MAMG-N), discussing also their particular sub-cases. The field equations of the theories are obtained by regarding the metric tensor and the general affine connection as independent variables. We then focus on the case in which the function characterizing the aforementioned metric-affine models is linear and consider a Friedmann-Lema\^{i}tre-Robertson-Walker background to study cosmological aspects and applications.

\vspace{1cm}


\vfill
\noindent {\small{\it
    E-mail:  \\
{\tt nmyrzakulov@gmail.com}; \\    
{\tt rmyrzakulov@gmail.com}; \\
{\tt lucrezia.ravera@polito.it}.}}
   \eject
	
\tableofcontents

\noindent\hrulefill

\section{Introduction}\label{intro}

In the 20th century, physics experienced extraordinary progress with the formulation of General Relativity (GR), Einstein's well-celebrated theory of gravity. But despite its great success and solid predictive power, GR is not devoid of limitations, which manifest as shortages at both very small and large scales \cite{Capozziello:2014yqa, Will:2014kxa}, and contradictions between theory and observations. In particular, the flaws of GR at large scales are grievous given that gravity appears to be the force that rules cosmic evolution. In this context, GR is indeed unable to explain the observed late time accelerated expansion of the universe. Another problematic point is the inability of GR to explicate the rotational curves of galaxies without the need of dark matter. Further open issues which highly involve gravity at cosmological scales and are related with each other include the horizon and so-called flatness problem, cosmic inflation and early universe, size, origin and future of the latter, the abundant and mysterious dark energy, the cosmological constant and coincidence problems.

As a consequence, many physicists will agree that gravity, even though related with phenomena that we experience in everyday life, still stands as the most enigmatic of the fundamental interactions. The lack of a clear understanding of gravity has led over the years to the formulation of disparate alternative theoretical frameworks, which collectively go by the name of modified gravity \cite{CANTATA:2021ktz}. In fact, the terminologies ``alternative theory of gravity'' and ``modified gravity'' have become standard for gravitational theories differing from the most conventional one, where the latter is considered to be GR (whose rigorous mathematical formulations resides in Riemannian geometry). The litterature on the subject is huge. Let us mention, for instance, (Palatini and metric-affine) $f(R)$ gravity \cite{Sotiriou:2008rp,Iosifidis:2018zjj,Capozziello:2009mq,Klemm:2020mfp,Klemm:2020gfm}, the teleparallel $f(T)$ gravity theories \cite{Aldrovandi:2021,Myrzakulov:2010vz}, symmetric teleparallel $f(Q)$ gravity \cite{Nester:1998mp,BeltranJimenez:2018vdo}, and scalar-tensor theories \cite{Bartolo:1999sq,Charmousis:2012}. Here, let us mention that teleparallel theories of gravity are typically ``gauge'' theories of gravity.\footnote{We put the word ``gauge'' in inverted commas as, commonly, gravitational theories are invariant under diffeomorphisms by construction, but they are not invariant under spacetime translations. Thus, they are not true gauge theory of the associated group considered. However, we shall adopt the terminology of ``gauge theory of gravity'' since it is widely used in the literature, keeping in mind that, in fact, in that cases we just have diffeomorphisms invariance rather than invariance under spacetime translations.} In particular, one can either fix a priori symmetries and form of the connection in such a way to determine the explicit form of torsion (cf., e.g., \cite{Cai:2015emx}) or non-metricity in terms of what turns out to be the dynamical field (i.e., in the tetrads formalism, the vielbein), or start with no a priori assumption on the symmetries of the connection which, however, results to be related to non-Riemannian quantities by the definition of the field-strengths associated with the gauge group at hand. Differently, in this review we will consider gravitational theories out of this gauge realm, as we will further discuss in the following.

In the context of modified gravity, a wide part of the physics scientific community claims that the understanding and solution to open issues regarding the gravitational interaction may need generalizations and extensions of Riemannian geometry. 
It is well-known that one way to go beyond Riemannian geometry is to release the Riemannian assumptions of metric compatibility and torsionlessness of the connection and therefore allow, as we have already anticipated above, for non-vanishing torsion and non-metricity (along with curvature). This is the framework of non-Riemannian geometry\footnote{Regarding non-Riemannian geometry, we refer the reader to \cite{Eisenhart:1927,Schouten:2013}. Besides, we highlight \cite{Klemm:2018bil} for a recent review of Einstein manifolds with torsion and non-metricity and applications in physics (for further interesting applications cf., for instance, \cite{Klemm:2019izb} and references therein).} and, in particular, the ``geometric arena'' where Metric-Affine Gravity (MAG) theories are developed \cite{Hehl:1994ue,Vitagliano:2010sr,Iosifidis:2019jgi,Percacci:2020bzf}. In particular, even if the metric-affine approach has been widely used in order to interpret gravity as a gauge theory, there is no conceptual or physical problem in studying metric-affine theories outside this
realm. This is in fact what we are going do in the present paper, where we will not deal with gauge theories of gravity. Specifically, we will work in the first order formalism, considering the metric and the connection a priori as independent, without assuming any symmetry or constraint on the connection from the very beginning. In this setting, the final form of the connection in terms of non-Riemannian objects is then obtained from the study of the field equations of the theory.

MAGs in the first order formalism are gravitational theories alternative to GR exhibiting a very general setup, with the potential of properly describing various physical scenarios, where the metric and the general affine connection (i.e., involving, in principle, torsion and non-metricity) are considered, a priori, as independent. Especially, no symmetry is imposed a priori on the connection. An additional motivation for studying MAG theories emerges when one considers coupling with matter, as the matter Lagrangian depends on the connection as well. Therefore, in MAG there is a new physical object that comes into play when varying the matter part of the action with respect to the connection, that is the so-called hypermomentum tensor \cite{Hehl:1976,Hehl:1976kt,Hehl:1976kv}, which encompasses the microscopic characteristics of matter. In this setup, the energy-momentum tensor sources spacetime curvature by means of the metric field equations, while the hypermomentum is source of spacetime torsion and non-metricity through the connection field equations.

Following the line of thought based on the idea that considering alternative geometric frameworks one can effectively gain better insights towards a deeper and more complete understanding of gravity than the one provided by GR, in this paper we collect and review a rather general class of gravity theories, the so-called Myrzakulov Gravity (MG) models, also referred to as MG-N, $\rm{N=I,II, \ldots, VIII}$ \cite{Myrzakulov:2012ug}, and derive their respective metric-affine generalizations, which will go by the name of Metric-Affine Myrzakulov Gravity (MAMG) models. The action of the MG theories is characterized by a generic function $F$ of non-Riemannian scalars (the scalar curvature of the general affine connection, the torsion scalar, the non-metricity scalar, and, besides, the energy-momentum trace), which takes different form depending on the specific model. Moreover, in the metric-affine framework one may generalize the theories by including a dependence on the divergence of the dilation current, the latter being a trace of the hypermomentum tensor, in $F$.
We will consider four spacetime dimensions and work in the first order (Palatini) formalism, where the metric tensor $g_{\mu \nu}$ and the general affine connection ${\Gamma^\lambda}_{\mu \nu}$ are treated, a priori, as independent variables, following the lines of \cite{Iosifidis:2021kqo}.
Subsequently, we focus on the case in which the function characterizing the aforementioned metric-affine models is linear and consider a Friedmann-Lema\^{i}tre-Robertson-Walker (FLRW) background to study cosmological aspects.

The paper is organized as follows: In Sections \ref{mamg1} to \ref{mamg8} we review the MG-N models and present their respective metric-affine generalizations (MAMG-N), together with their particular sub-cases, while Section \ref{cosmolasp} is devoted to the study of a non-Riemannian cosmological setup in which, in particular, we derive the modified Friedmann equations for the linear MAMG-N theories and discuss cosmological applications of the results (another application is also presented in Appendix \ref{FTplusscal}). In Section \ref{concl} we make some final remark and discuss possible future developments. In Appendix \ref{appa} we collect notation, conventions, definitions, and useful MAG formulas, while Appendix \ref{appb} gathers key expressions in the context of cosmology with torsion and non-metricity in a homogeneous, non-Riemannian FLRW spacetime.

\section{MG-I and MAMG-I}\label{mamg1}

We first describe the MG-I model. Its action reads \cite{Myrzakulov:2012ug}
\begin{equation}\label{MG1}
\mathcal{S}^{(\rm{I})}[g,\Gamma,\varphi] = \frac{1}{2 \kappa} \int \sqrt{-g} d^4 x \left[ F(R,T) + 2 \kappa \mathcal{L}_{\text{m}} \right] \,,
\end{equation}
where $\kappa=8 \pi G$ is the gravitational constant, $R$ is the curvature scalar of the general affine connection ${\Gamma^\lambda}_{\mu \nu}$ involving torsion and non-metricity and $T$ is the torsion scalar (see Appendix \ref{appa}). In eq. \eqref{MG1}, $F=F(R,T)$ is a generic function of $R$ and $T$. In fact, the MG-I model represents an extension of both the $F(R)$ and $F(T)$ gravity theories. The action \eqref{MG1} depends on the metric field $g_{\mu \nu}$, the affine connection ${\Gamma^\lambda}_{\mu \nu}$, and the matter fields, collectively denoted by $\varphi$, appearing in the matter Lagrangian $\mathcal{L}_{\text{m}}$. \\
The variation of \eqref{MG1} with respect to the metric field yields
\begin{equation}\label{deltagMG1}
- \frac{1}{2} g_{\mu \nu} F + F'_R R_{(\mu \nu)} + F'_T \left( 2 S_{\nu \alpha \beta} {S_\mu}^{\alpha \beta} - S_{\alpha \beta \mu} {S^{\alpha \beta}}_\nu + 2 S_{\nu \alpha \beta} {S_\mu}^{\beta \alpha} - 4 S_\mu S_\nu \right) = \kappa T_{\mu \nu} \,,
\end{equation}
where $R_{(\mu \nu)}$ is the symmetric part of the Ricci tensor of $\Gamma$, ${S_{\mu \nu}}^\lambda$ is the torsion tensor, $S_\mu$ is the torsion trace, $T_{\mu \nu}$ is the energy-momentum tensor (cf. the respective definitions in Appendix \ref{appa}), and $F'_R:=\frac{dF}{dR}$, $F'_T:=\frac{dF}{dT}$.\footnote{Here and in the following we adopt the notation $F'_X:=\frac{dF}{dX}$ to denote the derivative of $F$ with respect to any scalar $X$ of which $F$ is function.} \\
On the other hand, the connection field equations are
\begin{equation}\label{deltagammaMG1}
{P_\lambda}^{\mu \nu} (F'_R) + 2 F'_T \left({S^{\mu \nu}}_\lambda - 2 {S_\lambda}^{[\mu \nu]} - 4 S^{[\mu} \delta^{\nu]}_\lambda \right) = 0 \,,
\end{equation}
where ${P_\lambda}^{\mu \nu} (F'_R)$ is the modified Palatini tensor,
\begin{equation}\label{modpala}
{P_\lambda}^{\mu \nu} (F'_R) := - \frac{\nabla_\lambda \left(\sqrt{-g} F'_R g^{\mu \nu} \right)}{\sqrt{-g}} + \frac{\nabla_\alpha \left( \sqrt{-g} F'_R g^{\mu \alpha}\delta^\nu_\lambda \right)}{\sqrt{-g}} + 2 F'_R \left( S_\lambda g^{\mu \nu} - S^\mu \delta^\nu_\lambda - {S_\lambda}^{\mu \nu} \right) \,,
\end{equation}
being $\nabla$ the covariant derivative associated with the general affine connection $\Gamma$.

\subsection{Metric-affine generalizations of the MG-I model}

As we have already mentioned in the introduction, in the metric-affine setup the matter Lagrangian depends on the connection as well. In this framework, the theory is assumed to have, in principle, a non-vanishing hypermomentum tensor, ${\Delta_\lambda}^{\mu \nu} := - \frac{2}{\sqrt{-g}} \frac{\delta \left( \sqrt{-g} \mathcal{L}_{\text{m}}\right)}{\delta {\Gamma^\lambda}_{\mu \nu}}$, which is one of the sources of MAG theories (along with the energy-momentum tensor, cf. Appendix \ref{appa}). The hypermomentum has a direct physical interpretation when split into its irreducible pieces of spin, dilation, and shear \cite{Hehl:1976,Hehl:1976kt,Hehl:1976kv}. Besides, as observed in \cite{Iosifidis:2021kqo}, in the metric-affine setup one may also consider the function $F$ appearing in the various MG models to depend on the contribution 
\begin{equation}\label{diverghyperm}
\mathcal{D} := \frac{1}{\sqrt{-g}} \partial_\nu \left( \sqrt{-g} \Delta^\nu \right) \,,
\end{equation}
where
\begin{equation}
\Delta^\nu := {\Delta_\mu}^{\mu \nu} 
\end{equation}
is the dilation current.\footnote{In particular, the energy-momentum trace $\mathcal{T}:=g^{\mu \nu}T_{\mu \nu}$ and $\mathcal{D}$ can be placed on an equal footing (see \cite{Iosifidis:2021kqo} for details). The first, in fact, will appear in the MG-IV, MG-VI, MG-VII, and MG-VIII theories and in their metric-affine generalization, in which case we will also include the $\mathcal{D}$ contribution.} \\
Taking all of this into account, the MAMG-I action, that is the metric-affine generalization of \eqref{MG1}, reads as follows:
\begin{equation}\label{aMAMG1}
\mathcal{S}^{(\rm{I})}_{\text{MAMG}} = \frac{1}{2 \kappa} \int \sqrt{-g} d^4 x \left[ F(R,T,\mathcal{D}) + 2 \kappa \mathcal{L}_{\text{m}} \right] \,,
\end{equation}
where we have also introduced a dependence on $\mathcal{D}$ in the function $F$. \\
The metric field equations of the theory are
\begin{equation}\label{deltagMAMG1}
- \frac{1}{2} g_{\mu \nu} F + F'_R R_{(\mu \nu)} + F'_T \left( 2 S_{\nu \alpha \beta} {S_\mu}^{\alpha \beta} - S_{\alpha \beta \mu} {S^{\alpha \beta}}_\nu + 2 S_{\nu \alpha \beta} {S_\mu}^{\beta \alpha} - 4 S_\mu S_\nu \right) + F'_{\mathcal{D}} M_{\mu \nu} = \kappa T_{\mu \nu} \,,
\end{equation}
where
\begin{equation}\label{Mmunu}
M_{\mu \nu} := \frac{\delta \mathcal{D}}{\delta g^{\mu \nu}} \,,
\end{equation}
while the variation of \eqref{aMAMG1} with respect to the general affine connection ${\Gamma^\lambda}_{\mu \nu}$ yields
\begin{equation}\label{deltagammaMAMG1}
{P_\lambda}^{\mu \nu} (F'_R) + 2 F'_T \left({S^{\mu \nu}}_\lambda - 2 {S_\lambda}^{[\mu \nu]} - 4 S^{[\mu} \delta^{\nu]}_\lambda \right) - {M_\lambda}^{\mu \nu \rho} \partial_\rho F'_{\mathcal{D}} = \kappa {\Delta_\lambda}^{\mu \nu} \,,
\end{equation}
where we have defined
\begin{equation}\label{Mlmnr}
{M_\lambda}^{\mu \nu \rho} := \frac{\delta \Delta^\rho}{\delta {\Gamma^\lambda}_{\mu \nu}} \,.
\end{equation}
Let us mention, here, that one could also consider a ``minimal'' metric-affine generalization of the MG-I theory, obtained without including a dependence on $\mathcal{D}$ in the function $F$. In this case, the metric field equations would coincide with those of the MG-I model, namely \eqref{deltagMG1}, while the connection field equations would be
\begin{equation}\label{mindeltagammaMAMG1}
{P_\lambda}^{\mu \nu} (F'_R) + 2 F'_T \left({S^{\mu \nu}}_\lambda - 2 {S_\lambda}^{[\mu \nu]} - 4 S^{[\mu} \delta^{\nu]}_\lambda \right) = \kappa {\Delta_\lambda}^{\mu \nu} \,,
\end{equation}
as we still have the hypermomentum contribution obtained by varying the matter Lagrangian $\mathcal{L}_{\text{m}}$ with respect to the connection.
The same result is obtained for specific matter such that $\Delta^\nu=0$.\footnote{In fact, for specific matter one has an explicit, specific expression for the whole ${\Delta_\lambda}^{\mu \nu}$.}

\subsection{Particular sub-cases of the MAMG-I theory}

In this subsection we collect some particular sub-cases of the MAMG-I model. In each sub-case, if we remove the dependence of the matter Lagrangian on the general affine connection, we are left with the respective sub-case of the MG-I model. We have already mentioned above one sub-case, corresponding to the minimal MAMG-I theory (minimal metric-affine generalization of MG-I), that is the metric-affine $F(R,T)$ theory. Let us discuss other sub-cases in the following.

\subsubsection{Metric-affine $F(R)$ theory}\label{frgrav}

First, we restrict ourselves to $F(R,T,\mathcal{D}) \to F(R)$, that is we assume that $F$ is independent of the torsion scalar $T$ and we also remove the $\mathcal{D}$ dependence in $F$. Then, the MAMG-I model reduces to the well-known metric-affine $F(R)$ gravity. The action of the theory has the form
\begin{equation}\label{fract}
\mathcal{S}_{F(R)} = \frac{1}{2 \kappa} \int \sqrt{-g} d^4 x \left[ F(R) + 2 \kappa \mathcal{L}_{\text{m}} \right] \,,
\end{equation}
where $R$ is the curvature scalar of the general affine connection. The variation of the action \eqref{fract} with respect to the metric and the affine connection gives the following set of field equations:
\begin{equation}
\begin{split}
& - \frac{1}{2} g_{\mu \nu} F + F'_R R_{(\mu \nu)} = \kappa T_{\mu \nu} \,, \\
& {P_\lambda}^{\mu \nu} (F'_R) = \kappa {\Delta_\lambda}^{\mu \nu} \,,
\end{split}
\end{equation}
where ${P_\lambda}^{\mu \nu} (F'_R)$ is the modified Palatini tensor defined in \eqref{modpala}.

\subsubsection{Metric-affine $F(T)$ theory}\label{ftgrav}

Now we assume that $F(R,T,\mathcal{D}) \to F(T)$. This corresponds to the metric-affine $F(T)$ theory, which is another sub-case of the MAMG-I model. The action of the theory reads
\begin{equation}\label{ftact}
\mathcal{S}_{F(T)} = \frac{1}{2 \kappa} \int \sqrt{-g} d^4 x \left[ F(T) + 2 \kappa \mathcal{L}_{\text{m}} \right] \,,
\end{equation}
where $T$ is the torsion scalar. Varying the action \eqref{ftact} with respect to the metric field and the general affine connection, we get
\begin{equation}
\begin{split}
& - \frac{1}{2} g_{\mu \nu} F + F'_T \left( 2 S_{\nu \alpha \beta} {S_\mu}^{\alpha \beta} - S_{\alpha \beta \mu} {S^{\alpha \beta}}_\nu + 2 S_{\nu \alpha \beta} {S_\mu}^{\beta \alpha} - 4 S_\mu S_\nu \right) = \kappa T_{\mu \nu} \,, \\
& 2 F'_T \left({S^{\mu \nu}}_\lambda - 2 {S_\lambda}^{[\mu \nu]} - 4 S^{[\mu} \delta^{\nu]}_\lambda \right) = \kappa {\Delta_\lambda}^{\mu \nu} \,,
\end{split}
\end{equation}
respectively.
Here observe that, defining a specific form for the connection from the very beginning, as done in \cite{Cai:2015emx}, where the Weitzenb\"{o}ck connection \cite{Weitzenbock:1923} was considered, in $F(T)$ gravity one could then derive the Einstein equations for the metric field by exploiting the tetrads formalism, that is introducing the vielbein vector with the use of its component form $\textbf{V}_a=V^\mu_a \partial_\mu$, with $a=0,1,2,3$, and expressing the torsion in terms of the latter.\footnote{For a discussion on the degrees of freedom in $F(T)$ gravity we refer the reader to \cite{Cai:2015emx}.} Nevertheless, recall that here we are not assuming a specific form for the connection a priori and we are not dealing with a gauge theory of gravity. In our formalism, the final form of the connection arise when one studies in detail the field equations of the theory for given matter field contents, that is once the explicit expression of ${\Delta_\lambda}^{\mu \nu}$ is consequently found. Hence, torsion and, in particular, hypermomentum variables (i.e., sources) encode the full dynamics of the theory. Notice that analogous arguments apply to the $F(Q)$ (sub-)case we will briefly review in Section \ref{mamg2}.

\subsubsection{Metric-affine $F(R,\mathcal{D})$ theory}\label{frdgrav}

Let us now restrict ourselves to $F(R,T,\mathcal{D}) \to F(R,\mathcal{D})$, that is we assume that $F$ is independent of the torsion scalar $T$. In this case, the MAMG-I model boils down to the metric-affine $F(R,\mathcal{D})$ theory, which is an extension of the $F(R)$ gravity involving a dependence on the divergence of the dilation current $\mathcal{D}$ in $F$. The action of the model at hand has the form
\begin{equation}\label{frdact}
\mathcal{S}_{F(R,\mathcal{D})} = \frac{1}{2 \kappa} \int \sqrt{-g} d^4 x \left[ F(R,\mathcal{D}) + 2 \kappa \mathcal{L}_{\text{m}} \right] \,.
\end{equation}
The variation of the action \eqref{frdact} with respect to the metric and the general affine connection gives the following set of field equations:
\begin{equation}
\begin{split}
& - \frac{1}{2} g_{\mu \nu} F + F'_R R_{(\mu \nu)} + F'_{\mathcal{D}} M_{\mu \nu} = \kappa T_{\mu \nu} \,, \\
& {P_\lambda}^{\mu \nu} (F'_R) - {M_\lambda}^{\mu \nu \rho} \partial_\rho F'_{\mathcal{D}} = \kappa {\Delta_\lambda}^{\mu \nu} \,,
\end{split}
\end{equation}
with ${P_\lambda}^{\mu \nu} (F'_R)$ defined in \eqref{modpala}, while $M_{\mu \nu}$ and ${M_\lambda}^{\mu \nu \rho}$ are given by \eqref{Mmunu} and \eqref{Mlmnr}, respectively.

\subsubsection{Metric-affine $F(T,\mathcal{D})$ theory}\label{ftdgrav}

Considering $F(R,T,\mathcal{D}) \to F(T,\mathcal{D})$, the MAMG-I model reduces to the metric-affine $F(T,\mathcal{D})$ theory, which, in turn, is an extension of the $F(T)$ theory involving a dependence on the divergence of the dilation current in $F$. The action of the theory reads
\begin{equation}\label{ftdact}
\mathcal{S}_{F(T,\mathcal{D})} = \frac{1}{2 \kappa} \int \sqrt{-g} d^4 x \left[ F(T,\mathcal{D}) + 2 \kappa \mathcal{L}_{\text{m}} \right] \,,
\end{equation}
where $T$ is the torsion scalar and $\mathcal{D}$ the divergence of the dilation current. Varying the action \eqref{ftdact} with respect to the metric and the general affine connection, we find
\begin{equation}
\begin{split}
& - \frac{1}{2} g_{\mu \nu} F + F'_T \left( 2 S_{\nu \alpha \beta} {S_\mu}^{\alpha \beta} - S_{\alpha \beta \mu} {S^{\alpha \beta}}_\nu + 2 S_{\nu \alpha \beta} {S_\mu}^{\beta \alpha} - 4 S_\mu S_\nu \right) + F'_{\mathcal{D}} M_{\mu \nu} = \kappa T_{\mu \nu} \,, \\
& 2 F'_T \left({S^{\mu \nu}}_\lambda - 2 {S_\lambda}^{[\mu \nu]} - 4 S^{[\mu} \delta^{\nu]}_\lambda \right) - {M_\lambda}^{\mu \nu \rho} \partial_\rho F'_{\mathcal{D}} = \kappa {\Delta_\lambda}^{\mu \nu} \,,
\end{split}
\end{equation}
respectively.

\section{MG-II and MAMG-II}\label{mamg2}

In this section we start from the description of the MG-II theory and subsequently generalize it by considering a metric-affine setup.
The action of the MG-II model is \cite{Myrzakulov:2012ug}
\begin{equation}\label{MG2}
\mathcal{S}^{(\rm{II})}[g,\Gamma,\varphi] = \frac{1}{2 \kappa} \int \sqrt{-g} d^4 x \left[ F(R,Q) + 2 \kappa \mathcal{L}_{\text{m}} \right] \,.
\end{equation}
It is an extension of both the $F(R)$ and $F(Q)$ theories. Indeed, the function $F=F(R,Q)$ in \eqref{MG2} is a generic function of the scalar curvature $R$ (of the general affine connection $\Gamma$) and of $Q$, where $Q$ is the non-metricity scalar (cf. Appendix \ref{appa}). For some cosmological implications of the MG-II model we refer the reader to \cite{Myrzakulov:2012qp,Saridakis:2019qwt}, while observational constraints on the theory were studied in \cite{Anagnostopoulos:2020lec}. \\
The metric field equations of the theory read as follows:
\begin{equation}\label{deltagMG2}
- \frac{1}{2} g_{\mu \nu} F + F'_R R_{(\mu \nu)} + F'_Q L_{(\mu \nu)} + \hat{\nabla}_\lambda \left(F'_Q {J^\lambda}_{(\mu \nu)} \right) + g_{\mu \nu} \hat{\nabla}_\lambda \left(F'_Q \zeta^\lambda \right) = \kappa T_{\mu \nu} \,,
\end{equation}
having defined
\begin{equation}\label{hatder}
\hat{\nabla}_\lambda := \frac{1}{\sqrt{-g}} \left( 2 S_\lambda - \nabla_\lambda \right)
\end{equation}
and 
\begin{equation}\label{nonmetquant}
\begin{split}
L_{\mu \nu} & := \frac{1}{4} \left[ \left(Q_{\mu \alpha \beta} - 2 Q_{\alpha \beta \mu} \right) {Q_\nu}^{\alpha \beta} + \left( Q_\mu + 2 q_\mu \right) Q_\nu + \left( 2 Q_{\mu \nu \alpha} - Q_{\alpha \mu \nu} \right) Q^\alpha \right] \\
& \phantom{:= \,} - {\Xi^{\alpha \beta}}_\nu Q_{\alpha \beta \mu} - \Xi_{\alpha \mu \beta} {Q^{\alpha \beta}}_\nu \,, \\
{J^\lambda}_{\mu \nu} & := \sqrt{-g} \left(\frac{1}{4} {Q^\lambda}_{\mu \nu} - \frac{1}{2} {Q_{\mu \nu}}^\lambda + {\Xi^\lambda}_{\mu \nu} \right) \,, \\
\zeta^\lambda & := \sqrt{-g} \left( - \frac{1}{4} Q^\lambda + \frac{1}{2} q^\lambda \right) \,, 
\end{split}
\end{equation}
where $Q_{\lambda \mu \nu}$ is the non-metricity tensor, $Q_\lambda$ and $q_\lambda$ are its trace parts (see Appendix \ref{appa}), and $\Xi_{\lambda \mu \nu}$ is the so-called (non-metricity) ``superpotential'' defined in the first line of \eqref{superpot}. \\
The connection field equations are
\begin{equation}\label{deltagammaMG2}
{P_\lambda}^{\mu \nu} (F'_R) + F'_Q \left[ 2 {Q^{[\nu \mu]}}_\lambda - {Q_\lambda}^{\mu \nu} + \left( q^\nu - Q^\nu \right) \delta^\mu_\lambda + Q_\lambda g^{\mu \nu} + \frac{1}{2} Q^\mu \delta^\nu_\lambda \right] = 0 \,,
\end{equation}
where ${P_\lambda}^{\mu \nu} (F'_R)$ is given by \eqref{modpala}.

\subsection{Metric-affine generalizations of the MG-II model}

We can now move on to the metric-affine generalization of the MG-II theory.
Hence, taking into account the discussion in Section \ref{mamg1}, one may write the MAMG-II action as follows:
\begin{equation}\label{aMAMG2}
\mathcal{S}^{(\rm{II})}_{\text{MAMG}} = \frac{1}{2 \kappa} \int \sqrt{-g} d^4 x \left[ F(R,Q,\mathcal{D}) + 2 \kappa \mathcal{L}_{\text{m}} \right] \,,
\end{equation}
where, in particular, we have introduced in $F$ a dependence on $\mathcal{D}$, the latter being the divergence of the dilation current, defined in eq. \eqref{diverghyperm}. \\
The variation of \eqref{aMAMG2} with respect to the metric yields
\begin{equation}\label{deltagMAMG2}
- \frac{1}{2} g_{\mu \nu} F + F'_R R_{(\mu \nu)} + F'_Q L_{(\mu \nu)} + \hat{\nabla}_\lambda \left(F'_Q {J^\lambda}_{(\mu \nu)} \right) + g_{\mu \nu} \hat{\nabla}_\lambda \left(F'_Q \zeta^\lambda \right) + F'_{\mathcal{D}} M_{\mu \nu} = \kappa T_{\mu \nu} \,,
\end{equation}
where $M_{\mu \nu}$ has been defined in \eqref{Mmunu}. \\
On the other hand, the connection field equations of the MAMG-II theory read
\begin{equation}\label{deltagammaMAMG2}
{P_\lambda}^{\mu \nu} (F'_R) + F'_Q \left[ 2 {Q^{[\nu \mu]}}_\lambda - {Q_\lambda}^{\mu \nu} + \left( q^\nu - Q^\nu \right) \delta^\mu_\lambda + Q_\lambda g^{\mu \nu} + \frac{1}{2} Q^\mu \delta^\nu_\lambda \right] - {M_\lambda}^{\mu \nu \rho} \partial_\rho F'_{\mathcal{D}} = \kappa {\Delta_\lambda}^{\mu \nu} \,,
\end{equation}
where ${M_\lambda}^{\mu \nu \rho}$ is given by \eqref{Mlmnr}.
Analogously to what we have mentioned in Section \ref{mamg1} for the $\rm{N=I}$ case, one might also consider a minimal metric-affine generalization of the MG-II theory, excluding the $\mathcal{D}$ dependence in the function $F$. Then, the metric field equations would coincide with \eqref{deltagMG2}, while the connection field equations would be
\begin{equation}\label{mindeltagammaMAMG2}
{P_\lambda}^{\mu \nu} (F'_R) + F'_Q \left[ 2 {Q^{[\nu \mu]}}_\lambda - {Q_\lambda}^{\mu \nu} + \left( q^\nu - Q^\nu \right) \delta^\mu_\lambda + Q_\lambda g^{\mu \nu} + \frac{1}{2} Q^\mu \delta^\nu_\lambda \right] = \kappa {\Delta_\lambda}^{\mu \nu} \,.
\end{equation}
The same result can be obtained for specific matter couplings fulfilling $\Delta^\nu=0$.

\subsection{Particular sub-cases of the MAMG-II theory}

Here we collect some particular sub-cases of the MAMG-II theory. In each sub-case, if we remove the dependence of the matter Lagrangian on the general affine connection, we are left with the respective sub-case of the MG-II model. We have already mentioned above one sub-case, corresponding to the minimal MAMG-II model (minimal metric-affine generalization of MG-II), that is the metric-affine $F(R,Q)$ theory. Other two sub-cases correspond to the metric-affine $F(R)$ and $F(R,\mathcal{D})$ theories previously discussed (cf. \eqref{fract} and \eqref{frdact}, respectively). Let us report in the following also other two sub-cases.

\subsubsection{Metric-affine $F(Q)$ theory}\label{fqgrav}

We first restrict ourselves to $F(R,Q,\mathcal{D}) \to F(Q)$, that is we assume that $F$ is independent of the scalar curvature $R$ and we also remove the $\mathcal{D}$ dependence in $F$. With this assumption the MAMG-II model reduces to the metric-affine $F(Q)$ theory, whose action is
\begin{equation}\label{fqact}
\mathcal{S}_{F(Q)} = \frac{1}{2 \kappa} \int \sqrt{-g} d^4 x \left[ F(Q) + 2 \kappa \mathcal{L}_{\text{m}} \right] \,,
\end{equation}
where $Q$ is the non-metricity scalar. The variation of the action \eqref{fqact} with respect to the metric and the affine connection gives the following set of field equations:
\begin{equation}
\begin{split}
& - \frac{1}{2} g_{\mu \nu} F + F'_Q L_{(\mu \nu)} + \hat{\nabla}_\lambda \left(F'_Q {J^\lambda}_{(\mu \nu)} \right) + g_{\mu \nu} \hat{\nabla}_\lambda \left(F'_Q \zeta^\lambda \right) = \kappa T_{\mu \nu} \,, \\
& F'_Q \left[ 2 {Q^{[\nu \mu]}}_\lambda - {Q_\lambda}^{\mu \nu} + \left( q^\nu - Q^\nu \right) \delta^\mu_\lambda + Q_\lambda g^{\mu \nu} + \frac{1}{2} Q^\mu \delta^\nu_\lambda \right] = \kappa {\Delta_\lambda}^{\mu \nu} \,,
\end{split}
\end{equation}
with $L_{\mu \nu}$, ${J^\lambda}_{\mu \nu}$, and $\zeta^\lambda$ given by \eqref{nonmetquant}.
Here one could then apply arguments analogous to those made in the case of the $F(T)$ theory.

\subsubsection{Metric-affine $F(Q,\mathcal{D})$ theory}\label{fqdgrav}

On the other hand, considering $F(R,Q,\mathcal{D}) \to F(Q,\mathcal{D})$, the MAMG-II model boils down to the metric-affine $F(Q,\mathcal{D})$ theory, which, in turn, is an extension of the $F(Q)$ theory above involving a dependence on the divergence of the dilation current in $F$. The action of the metric-affine $F(Q,\mathcal{D})$ model has the following form:
\begin{equation}\label{fqdact}
\mathcal{S}_{F(Q,\mathcal{D})} = \frac{1}{2 \kappa} \int \sqrt{-g} d^4 x \left[ F(Q,\mathcal{D}) + 2 \kappa \mathcal{L}_{\text{m}} \right] \,,
\end{equation}
where $Q$ is the non-metricity scalar and $\mathcal{D}$ the divergence of the dilation current. Varying the action \eqref{fqdact} with respect to the metric and the general affine connection, we obtain the set of field equations
\begin{equation}
\begin{split}
& - \frac{1}{2} g_{\mu \nu} F + F'_Q L_{(\mu \nu)} + \hat{\nabla}_\lambda \left(F'_Q {J^\lambda}_{(\mu \nu)} \right) + g_{\mu \nu} \hat{\nabla}_\lambda \left(F'_Q \zeta^\lambda \right) + F'_{\mathcal{D}} M_{\mu \nu} = \kappa T_{\mu \nu} \,, \\
& F'_Q \left[ 2 {Q^{[\nu \mu]}}_\lambda - {Q_\lambda}^{\mu \nu} + \left( q^\nu - Q^\nu \right) \delta^\mu_\lambda + Q_\lambda g^{\mu \nu} + \frac{1}{2} Q^\mu \delta^\nu_\lambda \right] - {M_\lambda}^{\mu \nu \rho} \partial_\rho F'_{\mathcal{D}} = \kappa {\Delta_\lambda}^{\mu \nu} \,,
\end{split}
\end{equation}
with $M_{\mu \nu}$ and ${M_\lambda}^{\mu \nu \rho}$ given by \eqref{Mmunu} and \eqref{Mlmnr}, respectively.

\section{MG-III and MAMG-III}\label{mamg3}

Let us start by describing the MG-III theory, whose action is \cite{Myrzakulov:2012ug}
\begin{equation}\label{MG3}
\mathcal{S}^{(\rm{III})}[g,\Gamma,\varphi] = \frac{1}{2 \kappa} \int \sqrt{-g} d^4 x \left[ F(T,Q) + 2 \kappa \mathcal{L}_{\text{m}} \right] \,.
\end{equation}
In this model, the function $F=F(T,Q)$ is a generic function of the torsion scalar $T$ and the non-metricity scalar $Q$. The action \eqref{MG3} is an extension of both the $F(T)$ and $F(Q)$ theories. \\
Varying \eqref{MG3} with respect to the metric field we get
\begin{equation}\label{deltagMG3}
\begin{split}
& - \frac{1}{2} g_{\mu \nu} F + F'_T \left( 2 S_{\nu \alpha \beta} {S_\mu}^{\alpha \beta} - S_{\alpha \beta \mu} {S^{\alpha \beta}}_\nu + 2 S_{\nu \alpha \beta} {S_\mu}^{\beta \alpha} - 4 S_\mu S_\nu \right) + F'_Q L_{(\mu \nu)} \\
& + \hat{\nabla}_\lambda \left(F'_Q {J^\lambda}_{(\mu \nu)} \right) + g_{\mu \nu} \hat{\nabla}_\lambda \left(F'_Q \zeta^\lambda \right) = \kappa T_{\mu \nu} \,,
\end{split}
\end{equation}
where $\hat{\nabla}$ is given by \eqref{hatder}, while the tensor $L_{\mu \nu}$ and the tensor and vector densities ${J^\lambda}_{\mu \nu}$ and $\zeta^\lambda$, respectively, are defined in \eqref{nonmetquant}. \\
On the other hand, from the variation of \eqref{MG3} with respect to the general affine connection ${\Gamma^\lambda}_{\mu \nu}$ we obtain
\begin{equation}\label{deltagammaMG3}
2 F'_T \left({S^{\mu \nu}}_\lambda - 2 {S_\lambda}^{[\mu \nu]} - 4 S^{[\mu} \delta^{\nu]}_\lambda \right) + F'_Q \left[ 2 {Q^{[\nu \mu]}}_\lambda - {Q_\lambda}^{\mu \nu} + \left( q^\nu - Q^\nu \right) \delta^\mu_\lambda + Q_\lambda g^{\mu \nu} + \frac{1}{2} Q^\mu \delta^\nu_\lambda \right] = 0 \,.
\end{equation}
We can now move on to the metric-affine generalization of the theory.

\subsection{Metric-affine generalizations of the MG-III model}

The action of the MAMG-III model is
\begin{equation}\label{aMAMG3}
\mathcal{S}^{(\rm{III})}_{\text{MAMG}} = \frac{1}{2 \kappa} \int \sqrt{-g} d^4 x \left[ F(T,Q,\mathcal{D}) + 2 \kappa \mathcal{L}_{\text{m}} \right] \,,
\end{equation}
where $\mathcal{D}$ is defined in \eqref{diverghyperm}. \\
The metric field equations of the theory read
\begin{equation}\label{deltagMAMG3}
\begin{split}
& - \frac{1}{2} g_{\mu \nu} F + F'_T \left( 2 S_{\nu \alpha \beta} {S_\mu}^{\alpha \beta} - S_{\alpha \beta \mu} {S^{\alpha \beta}}_\nu + 2 S_{\nu \alpha \beta} {S_\mu}^{\beta \alpha} - 4 S_\mu S_\nu \right) + F'_Q L_{(\mu \nu)} \\
& + \hat{\nabla}_\lambda \left(F'_Q {J^\lambda}_{(\mu \nu)} \right) + g_{\mu \nu} \hat{\nabla}_\lambda \left(F'_Q \zeta^\lambda \right) + F'_{\mathcal{D}} M_{\mu \nu} = \kappa T_{\mu \nu} \,,
\end{split}
\end{equation}
where $M_{\mu \nu}$ is given by \eqref{Mmunu}. \\
The connection field equations are
\begin{equation}\label{deltagammaMAMG3}
\begin{split}
& 2 F'_T \left({S^{\mu \nu}}_\lambda - 2 {S_\lambda}^{[\mu \nu]} - 4 S^{[\mu} \delta^{\nu]}_\lambda \right) + F'_Q \left[ 2 {Q^{[\nu \mu]}}_\lambda - {Q_\lambda}^{\mu \nu} + \left( q^\nu - Q^\nu \right) \delta^\mu_\lambda + Q_\lambda g^{\mu \nu} + \frac{1}{2} Q^\mu \delta^\nu_\lambda \right] \\
& - {M_\lambda}^{\mu \nu \rho} \partial_\rho F'_{\mathcal{D}} = \kappa {\Delta_\lambda}^{\mu \nu} \,,
\end{split}
\end{equation}
where ${M_\lambda}^{\mu \nu \rho}$ is defined in \eqref{Mlmnr}.
There also exists a minimal metric-affine generalization of the MG-III theory, which excludes the $\mathcal{D}$ dependence in the function $F$. In this minimal case, the metric field equations coincide with \eqref{deltagMG3}, while the connection field equations are
\begin{equation}\label{mindeltagammaMAMG3}
\begin{split}
& 2 F'_T \left({S^{\mu \nu}}_\lambda - 2 {S_\lambda}^{[\mu \nu]} - 4 S^{[\mu} \delta^{\nu]}_\lambda \right) + F'_Q \left[ 2 {Q^{[\nu \mu]}}_\lambda - {Q_\lambda}^{\mu \nu} + \left( q^\nu - Q^\nu \right) \delta^\mu_\lambda + Q_\lambda g^{\mu \nu} + \frac{1}{2} Q^\mu \delta^\nu_\lambda \right] \\
& = \kappa {\Delta_\lambda}^{\mu \nu} \,.
\end{split}
\end{equation}
The same equations can be obtained considering some specific matter couplings, always in the metric-affine framework, such that $\Delta^\nu=0$.

Let us conclude this section by mentioning that there are other particular sub-cases, besides the metric-affine $F(T,Q)$ one above, of the MAMG-III model, corresponding to the metric-affine $F(T)$, $F(T,\mathcal{D})$, $F(Q)$, and $F(Q,\mathcal{D})$ theories previously discussed (cf., respectively, \eqref{ftact}, \eqref{ftdact}, \eqref{fqact}, and \eqref{fqdact}).\footnote{In each sub-case, removing the dependence of the matter Lagrangian on the general affine connection, one is left with the respective sub-case of the MG-III model.}

\section{MG-IV and MAMG-IV}\label{mamg4}

We first describe the MG-IV theory, whose action is \cite{Myrzakulov:2012ug}
\begin{equation}\label{MG4}
\mathcal{S}^{(\rm{IV})}[g,\Gamma,\varphi] = \frac{1}{2 \kappa} \int \sqrt{-g} d^4 x \left[ F(R,T,\mathcal{T}) + 2 \kappa \mathcal{L}_{\text{m}} \right] \,,
\end{equation}
where $F=F(R,T,\mathcal{T})$ is a generic function of the scalar curvature $R$, the torsion scalar $T$, and the energy-momentum trace $\mathcal{T}:=g^{\mu \nu} T_{\mu \nu}$. The MG-IV model extends the $F(R,\mathcal{T})$ (cf. \cite{Harko:2011kv}), $F(T)$, $F(R)$, and $F(T,\mathcal{T})$ (cf. \cite{Harko:2014aja}) theories. In particular, \eqref{MG4} represents an extension of the MG-I theory, as it also involves a dependence on $\mathcal{T}$ in $F$. \\
The metric field equations of \eqref{MG4} are
\begin{equation}\label{deltagMG4}
\begin{split}
& - \frac{1}{2} g_{\mu \nu} F + F'_R R_{(\mu \nu)} + F'_T \left( 2 S_{\nu \alpha \beta} {S_\mu}^{\alpha \beta} - S_{\alpha \beta \mu} {S^{\alpha \beta}}_\nu + 2 S_{\nu \alpha \beta} {S_\mu}^{\beta \alpha} - 4 S_\mu S_\nu \right) \\
& + F'_{\mathcal{T}} \left( \Theta_{\mu \nu} + T_{\mu \nu} \right)   = \kappa T_{\mu \nu} \,,
\end{split}
\end{equation}
where $\hat{\nabla}$ is defined in \eqref{hatder}, $L_{\mu \nu}$, ${J^\lambda}_{\mu \nu}$ and $\zeta^\lambda$ are given in \eqref{nonmetquant}, and
\begin{equation}\label{thetamunu}
\Theta_{\mu \nu} := g^{\alpha \beta} \frac{\delta T_{\alpha \beta}}{\delta g^{\mu \nu}} \,.
\end{equation}
The connection field equations read as follows:
\begin{equation}\label{deltagammaMG4}
{P_\lambda}^{\mu \nu} (F'_R) + 2 F'_T \left({S^{\mu \nu}}_\lambda - 2 {S_\lambda}^{[\mu \nu]} - 4 S^{[\mu} \delta^{\nu]}_\lambda \right) = 0 \,,
\end{equation}
where ${P_\lambda}^{\mu \nu} (F'_R)$ is defined in \eqref{modpala}.
Here, observe that \eqref{deltagammaMG4} coincides with the connection field equations \eqref{deltagammaMG1} of the MG-I theory. This is so due to the fact that in the MG theories matter does not couple to the connection and, therefore, the dependence on $\mathcal{T}$ in $F$ does not affect the connection field equations. Accordingly, we have no contribution to the connection field equations from the matter Lagrangian $\mathcal{L}_{\text{m}}$.

\subsection{Metric-affine generalizations of the MG-IV model}

Following the discussion in Section \ref{mamg1}, we write the MAMG-IV action as follows:
\begin{equation}\label{aMAMG4}
\mathcal{S}^{(\rm{IV})}_{\text{MAMG}} = \frac{1}{2 \kappa} \int \sqrt{-g} d^4 x \left[ F(R,T,\mathcal{T},\mathcal{D}) + 2 \kappa \mathcal{L}_{\text{m}} \right] \,,
\end{equation}
where $\mathcal{D}$ is the hypermomentum trace given by \eqref{diverghyperm}. In particular, here the energy-momentum trace $\mathcal{T}$ and the divergence of the dilation current $\mathcal{D}$ are placed on an equal footing (cf. \cite{Iosifidis:2021kqo}). \\
Varying the action \eqref{aMAMG4} with respect to the metric field we obtain
\begin{equation}\label{deltagMAMG4}
\begin{split}
& - \frac{1}{2} g_{\mu \nu} F + F'_R R_{(\mu \nu)} + F'_T \left( 2 S_{\nu \alpha \beta} {S_\mu}^{\alpha \beta} - S_{\alpha \beta \mu} {S^{\alpha \beta}}_\nu + 2 S_{\nu \alpha \beta} {S_\mu}^{\beta \alpha} - 4 S_\mu S_\nu \right) \\
& + F'_{\mathcal{T}} \left( \Theta_{\mu \nu} + T_{\mu \nu} \right) + F'_{\mathcal{D}} M_{\mu \nu}  = \kappa T_{\mu \nu} \,,
\end{split}
\end{equation}
with $M_{\mu \nu}$ and $\Theta_{\mu \nu}$ defined in \eqref{Mmunu} and \eqref{thetamunu}, respectively. \\
On the other hand, the variation of \eqref{aMAMG4} with respect to the general affine connection ${\Gamma^\lambda}_{\mu \nu}$ yields
\begin{equation}\label{deltagammaMAMG4}
{P_\lambda}^{\mu \nu} (F'_R) + 2 F'_T \left({S^{\mu \nu}}_\lambda - 2 {S_\lambda}^{[\mu \nu]} - 4 S^{[\mu} \delta^{\nu]}_\lambda \right) - {M_\lambda}^{\mu \nu \rho} \partial_\rho F'_{\mathcal{D}} = F'_{\mathcal{T}} {\Theta_\lambda}^{\mu \nu} + \kappa {\Delta_\lambda}^{\mu \nu} \,,
\end{equation}
where ${M_\lambda}^{\mu \nu \rho}$ is given by \eqref{Mlmnr} and
\begin{equation}\label{Thetalmn}
{\Theta_\lambda}^{\mu \nu} := - \frac{\delta \mathcal{T}}{\delta {\Gamma^\lambda}_{\mu \nu}} \,.
\end{equation}
The latter appears in \eqref{deltagammaMAMG4} since, in the metric-affine setup, the energy-momentum trace $\mathcal{T}$ may have, in principle, a non-trivial dependence on the hypermomentum tensor.
Note that if matter does not couple to the connection (which is the case, for instance, of a classical perfect fluid with no inner structure) one has ${\Theta_\lambda}^{\mu \nu}=0$, ${\Delta_\lambda}^{\mu \nu}=0$, and ${M_\lambda}^{\mu \nu \rho}=0$ (together with $\mathcal{D}=0$, that is no hypermomentum contribution also in the metric field equations).
Let us mention that one can also consider a minimal metric-affine generalization of the MG-IV theory, excluding the $\mathcal{D}$ dependence in the function $F$. The metric field equations, in this case, would coincide with \eqref{deltagMG4}, while the connection field equations would be
\begin{equation}\label{mindeltagammaMAMG4}
{P_\lambda}^{\mu \nu} (F'_R) + 2 F'_T \left({S^{\mu \nu}}_\lambda - 2 {S_\lambda}^{[\mu \nu]} - 4 S^{[\mu} \delta^{\nu]}_\lambda \right) = F'_{\mathcal{T}} {\Theta_\lambda}^{\mu \nu} + \kappa {\Delta_\lambda}^{\mu \nu} \,.
\end{equation}
The same result would follow from some specific matter fulfilling $\Delta^\nu=0$.
Observe that if ${\Theta_\lambda}^{\mu \nu}=0$ the connection field equations \eqref{deltagammaMAMG4} reduce to \eqref{deltagammaMAMG1}, which are those of the MAMG-I theory. Analogously, if ${\Theta_\lambda}^{\mu \nu}=0$, then \eqref{mindeltagammaMAMG4} boil down to \eqref{mindeltagammaMAMG1}, namely to the connection field equations of the minimal MAMG-I model.

\subsection{Particular sub-cases of the MAMG-IV theory}

Regarding the MAMG-IV model, here we discuss its sub-cases. Two of them are the minimal MAMG-IV model, minimal metric-affine generalization of MG-IV, that is the metric-affine $F(R,T,\mathcal{T})$ theory, and, of course, the minimal MAMG-I model, minimal metric-affine generalization of MG-I, that is the metric-affine $F(R,T)$ theory. Other sub-cases consist of the $F(R)$, $F(T)$, $F(R,\mathcal{D})$, and $F(T,\mathcal{D})$ theories we described earlier (cf. \eqref{fract}, \eqref{ftact}, \eqref{frdact}, and \eqref{ftdact}, respectively). Let us briefly sketch other four sub-cases in the following. In each of the MAMG-IV sub-cases, if we remove the dependence of the matter Lagrangian on the general affine connection, we are left with the respective sub-cases of the MG-IV model.

\subsubsection{Metric-affine $F(R,\mathcal{T})$ theory}\label{frmtgrav}

Let us consider $F(R,T,\mathcal{T},\mathcal{D}) \to F(R,\mathcal{T})$. Then, the MAMG-IV model reduces to the metric-affine $F(R,\mathcal{T})$ theory \cite{Harko:2011kv}, that is also an extension of the metric-affine $F(R)$ gravity including a dependence on the energy-momentum trace $\mathcal{T}$ in $F$. The action of the theory has the form
\begin{equation}\label{frmtact}
\mathcal{S}_{F(R,\mathcal{T})} = \frac{1}{2 \kappa} \int \sqrt{-g} d^4 x \left[ F(R,\mathcal{T}) + 2 \kappa \mathcal{L}_{\text{m}} \right] \,,
\end{equation}
where $R$ is the curvature scalar of the general affine connection. The variation of the action \eqref{frmtact} with respect to the metric and the affine connection gives the following set of field equations:
\begin{equation}
\begin{split}
& - \frac{1}{2} g_{\mu \nu} F + F'_R R_{(\mu \nu)} + F'_{\mathcal{T}} \left( \Theta_{\mu \nu} + T_{\mu \nu} \right) = \kappa T_{\mu \nu} \,, \\
& {P_\lambda}^{\mu \nu} (F'_R) = F'_\mathcal{T} {\Theta_\lambda}^{\mu \nu} + \kappa {\Delta_\lambda}^{\mu \nu} \,,
\end{split}
\end{equation}
where ${P_\lambda}^{\mu \nu} (F'_R)$ is the modified Palatini tensor defined in \eqref{modpala}, while $\Theta_{\mu \nu}$ and ${\Theta_\lambda}^{\mu \nu}$ are given by eqs. \eqref{thetamunu} and \eqref{Thetalmn}, respectively.

\subsubsection{Metric-affine $F(T,\mathcal{T})$ theory}\label{ftmtgrav}

On the other hand, considering $F(R,T,\mathcal{T},\mathcal{D}) \to F(T,\mathcal{T})$, the MAMG-IV model boils down to the metric-affine $F(T,\mathcal{T})$ theory \cite{Harko:2014aja}, which is also an extension of the metric-affine $F(T)$ model as it includes a dependence on the energy-momentum trace $\mathcal{T}$ in $F$ as well. The action of the metric-affine $F(T,\mathcal{T})$ theory reads
\begin{equation}\label{ftmtact}
\mathcal{S}_{F(T,\mathcal{T})} = \frac{1}{2 \kappa} \int \sqrt{-g} d^4 x \left[ F(T,\mathcal{T}) + 2 \kappa \mathcal{L}_{\text{m}} \right] \,,
\end{equation}
where $T$ is the torsion scalar. From the variation of the action \eqref{ftmtact} with respect to the metric and the general affine connection we obtain
\begin{equation}
\begin{split}
& - \frac{1}{2} g_{\mu \nu} F + F'_T \left( 2 S_{\nu \alpha \beta} {S_\mu}^{\alpha \beta} - S_{\alpha \beta \mu} {S^{\alpha \beta}}_\nu + 2 S_{\nu \alpha \beta} {S_\mu}^{\beta \alpha} - 4 S_\mu S_\nu \right) + F'_{\mathcal{T}} \left( \Theta_{\mu \nu} + T_{\mu \nu} \right) = \kappa T_{\mu \nu} \,, \\
& 2 F'_T \left({S^{\mu \nu}}_\lambda - 2 {S_\lambda}^{[\mu \nu]} - 4 S^{[\mu} \delta^{\nu]}_\lambda \right) = F'_\mathcal{T} {\Theta_\lambda}^{\mu \nu} + \kappa {\Delta_\lambda}^{\mu \nu} \,,
\end{split}
\end{equation}
respectively.

\subsubsection{Metric-affine $F(R,\mathcal{T},\mathcal{D})$ theory}\label{frmtdgrav}

Now, we consider the sub-case in which $F(R,T,\mathcal{T},\mathcal{D}) \to F(R,\mathcal{T},\mathcal{D})$, that is we exclude only the $T$ dependence in $F$. In this case, the MAMG-IV theory reduces to the metric-affine $F(R,\mathcal{T},\mathcal{D})$ theory, which is an extension of the metric-affine $F(R,\mathcal{T})$ model \eqref{frmtact} with the inclusion of a dependence on the divergence of the dilation current in $F$. The action of the metric-affine $F(R,\mathcal{T},\mathcal{D})$ theory is given by
\begin{equation}\label{frmtdact}
\mathcal{S}_{F(R,\mathcal{T},\mathcal{D})} = \frac{1}{2 \kappa} \int \sqrt{-g} d^4 x \left[ F(R,\mathcal{T},\mathcal{D}) + 2 \kappa \mathcal{L}_{\text{m}} \right] \,.
\end{equation}
The variation of \eqref{frmtdact} with respect to the metric and the affine connection gives the following set of field equations:
\begin{equation}
\begin{split}
& - \frac{1}{2} g_{\mu \nu} F + F'_R R_{(\mu \nu)} + F'_{\mathcal{T}} \left( \Theta_{\mu \nu} + T_{\mu \nu} \right) + F'_{\mathcal{D}} M_{\mu \nu} = \kappa T_{\mu \nu} \,, \\
& {P_\lambda}^{\mu \nu} (F'_R) - {M_\lambda}^{\mu \nu \rho} \partial_\rho F'_{\mathcal{D}} = F'_\mathcal{T} {\Theta_\lambda}^{\mu \nu} + \kappa {\Delta_\lambda}^{\mu \nu} \,,
\end{split}
\end{equation}
with ${P_\lambda}^{\mu \nu} (F'_R)$ defined in \eqref{modpala}, while $M_{\mu \nu}$, ${M_\lambda}^{\mu \nu \rho}$, $\Theta_{\mu \nu}$, and ${\Theta_\lambda}^{\mu \nu}$ are given in eqs. \eqref{Mmunu}, \eqref{Mlmnr}, \eqref{thetamunu}, and \eqref{Thetalmn}, respectively.

\subsubsection{Metric-affine $F(T,\mathcal{T},\mathcal{D})$ theory}\label{ftmtdgrav}

Finally, taking $F(R,T,\mathcal{T},\mathcal{D}) \to F(T,\mathcal{T},\mathcal{D})$, the MAMG-IV theory reduces to the metric-affine $F(T,\mathcal{T},\mathcal{D})$ model, extension of the metric-affine $F(T,\mathcal{T})$ theory \eqref{ftmtact} with the inclusion of a dependence on the divergence of the dilation current in $F$. The action of the metric-affine $F(T,\mathcal{T},\mathcal{D})$ theory reads
\begin{equation}\label{ftmtdact}
\mathcal{S}_{F(T,\mathcal{T},\mathcal{D})} = \frac{1}{2 \kappa} \int \sqrt{-g} d^4 x \left[ F(T,\mathcal{T},\mathcal{D}) + 2 \kappa \mathcal{L}_{\text{m}} \right] \,.
\end{equation}
The variation of \eqref{ftmtdact} with respect to the metric and the general affine connection yields
\begin{equation}
\begin{split}
& - \frac{1}{2} g_{\mu \nu} F + F'_T \left( 2 S_{\nu \alpha \beta} {S_\mu}^{\alpha \beta} - S_{\alpha \beta \mu} {S^{\alpha \beta}}_\nu + 2 S_{\nu \alpha \beta} {S_\mu}^{\beta \alpha} - 4 S_\mu S_\nu \right) + F'_{\mathcal{T}} \left( \Theta_{\mu \nu} + T_{\mu \nu} \right) + F'_{\mathcal{D}} M_{\mu \nu} \\
& = \kappa T_{\mu \nu} \,, \\
& 2 F'_T \left({S^{\mu \nu}}_\lambda - 2 {S_\lambda}^{[\mu \nu]} - 4 S^{[\mu} \delta^{\nu]}_\lambda \right) - {M_\lambda}^{\mu \nu \rho} \partial_\rho F'_{\mathcal{D}} = F'_\mathcal{T} {\Theta_\lambda}^{\mu \nu} + \kappa {\Delta_\lambda}^{\mu \nu} \,,
\end{split}
\end{equation}
respectively.

\section{MG-V and MAMG-V}\label{mamg5}

In this section we start from the description of the MG-V model and then generalize the latter in the metric-affine framework.
The MG-V action is given by \cite{Myrzakulov:2012ug}
\begin{equation}\label{MG5}
\mathcal{S}^{(\rm{V})}[g,\Gamma,\varphi] = \frac{1}{2 \kappa} \int \sqrt{-g} d^4 x \left[ F(R,T,Q) + 2 \kappa \mathcal{L}_{\text{m}} \right] \,.
\end{equation}
The MG-V model is an extension of the $F(R)$, $F(T)$, and $F(Q)$ theories. In particular, it represents an extension of the MG-I model involving also a dependence on $Q$ in $F$, and an extension of the MG-II theory as well, involving a dependence on $T$ in $F$. In fact, $F=F(R,T,Q)$ is a generic function of the scalar curvature $R$ of the general affine connection ${\Gamma^\lambda}_{\mu \nu}$, the torsion scalar $T$, and the non-metricity scalar $Q$. \\
The variation of \eqref{MG5} with respect to the metric field yields
\begin{equation}\label{deltagMG5}
\begin{split}
& - \frac{1}{2} g_{\mu \nu} F + F'_R R_{(\mu \nu)} + F'_T \left( 2 S_{\nu \alpha \beta} {S_\mu}^{\alpha \beta} - S_{\alpha \beta \mu} {S^{\alpha \beta}}_\nu + 2 S_{\nu \alpha \beta} {S_\mu}^{\beta \alpha} - 4 S_\mu S_\nu \right) + F'_Q L_{(\mu \nu)} \\ 
& + \hat{\nabla}_\lambda \left(F'_Q {J^\lambda}_{(\mu \nu)} \right) + g_{\mu \nu} \hat{\nabla}_\lambda \left(F'_Q \zeta^\lambda \right) = \kappa T_{\mu \nu} \,,
\end{split}
\end{equation}
where $\hat{\nabla}$ is defined in \eqref{hatder}, while the tensor $L_{\mu \nu}$ and the densities ${J^\lambda}_{\mu \nu}$ and $\zeta^\lambda$ are given by \eqref{nonmetquant}. \\
The equations obtained by varying the action \ref{MG5} with respect to the general affine connection ${\Gamma^\lambda}_{\mu \nu}$ read as follows:
\begin{equation}\label{deltagammaMG5}
\begin{split}
& {P_\lambda}^{\mu \nu} (F'_R) + 2 F'_T \left({S^{\mu \nu}}_\lambda - 2 {S_\lambda}^{[\mu \nu]} - 4 S^{[\mu} \delta^{\nu]}_\lambda \right) \\
& + F'_Q \left[ 2 {Q^{[\nu \mu]}}_\lambda - {Q_\lambda}^{\mu \nu} + \left( q^\nu - Q^\nu \right) \delta^\mu_\lambda + Q_\lambda g^{\mu \nu} + \frac{1}{2} Q^\mu \delta^\nu_\lambda \right] = 0 \,,
\end{split}
\end{equation}
where ${P_\lambda}^{\mu \nu} (F'_R)$ is the modified Palatini tensor defined in \eqref{modpala}.

\subsection{Metric-affine generalizations of the MG-V model}

We can now move on to the metric-affine generalization of the MG-V theory.
Hence, in accordance with the discussion in Section \ref{mamg1}, let us write the MAMG-V action as
\begin{equation}\label{aMAMG5}
\mathcal{S}^{(\rm{V})}_{\text{MAMG}} = \frac{1}{2 \kappa} \int \sqrt{-g} d^4 x \left[ F(R,T,Q,\mathcal{D}) + 2 \kappa \mathcal{L}_{\text{m}} \right] \,,
\end{equation}
where $\mathcal{D}$ is defined in \eqref{diverghyperm}. \\
The variation of \eqref{aMAMG5} with respect to the metric yields
\begin{equation}\label{deltagMAMG5}
\begin{split}
& - \frac{1}{2} g_{\mu \nu} F + F'_R R_{(\mu \nu)} + F'_T \left( 2 S_{\nu \alpha \beta} {S_\mu}^{\alpha \beta} - S_{\alpha \beta \mu} {S^{\alpha \beta}}_\nu + 2 S_{\nu \alpha \beta} {S_\mu}^{\beta \alpha} - 4 S_\mu S_\nu \right) + F'_Q L_{(\mu \nu)} \\ 
& + \hat{\nabla}_\lambda \left(F'_Q {J^\lambda}_{(\mu \nu)} \right) + g_{\mu \nu} \hat{\nabla}_\lambda \left(F'_Q \zeta^\lambda \right) + F'_{\mathcal{D}} M_{\mu \nu} = \kappa T_{\mu \nu} \,,
\end{split}
\end{equation}
where $M_{\mu \nu}$ is given by \eqref{Mmunu}. \\
The connection field equations of the theory read
\begin{equation}\label{deltagammaMAMG5}
\begin{split}
& {P_\lambda}^{\mu \nu} (F'_R) + 2 F'_T \left({S^{\mu \nu}}_\lambda - 2 {S_\lambda}^{[\mu \nu]} - 4 S^{[\mu} \delta^{\nu]}_\lambda \right) \\
& + F'_Q \left[ 2 {Q^{[\nu \mu]}}_\lambda - {Q_\lambda}^{\mu \nu} + \left( q^\nu - Q^\nu \right) \delta^\mu_\lambda + Q_\lambda g^{\mu \nu} + \frac{1}{2} Q^\mu \delta^\nu_\lambda \right] - {M_\lambda}^{\mu \nu \rho} \partial_\rho F'_{\mathcal{D}} = \kappa {\Delta_\lambda}^{\mu \nu} \,,
\end{split}
\end{equation}
with ${M_\lambda}^{\mu \nu \rho}$ defined in \eqref{Mlmnr}.
Analogously to what we have seen for the previous models, here we mention that a minimal metric-affine generalization of the MG-V theory might also be considered, in which the function $F$ does not exhibits a dependence on the divergence of the dilation current $\mathcal{D}$. In the minimal case the metric field equations boil down to \eqref{deltagMG5} and the connection field equations become
\begin{equation}\label{mindeltagammaMAMG5}
\begin{split}
& {P_\lambda}^{\mu \nu} (F'_R) + 2 F'_T \left({S^{\mu \nu}}_\lambda - 2 {S_\lambda}^{[\mu \nu]} - 4 S^{[\mu} \delta^{\nu]}_\lambda \right) \\
& + F'_Q \left[ 2 {Q^{[\nu \mu]}}_\lambda - {Q_\lambda}^{\mu \nu} + \left( q^\nu - Q^\nu \right) \delta^\mu_\lambda + Q_\lambda g^{\mu \nu} + \frac{1}{2} Q^\mu \delta^\nu_\lambda \right] = \kappa {\Delta_\lambda}^{\mu \nu} \,,
\end{split}
\end{equation}
One can end up with the same equations by considering some specific matter coupling, always in the metric-affine setup, satisfying $\Delta^\nu=0$.

Always concerning the MAMG-V theory, besides the sub-case we have already mentioned above, corresponding to the minimal MAMG-V model (minimal metric-affine generalization of MG-V, that is the metric-affine $F(R,T,Q)$ theory), we also have the following sub-cases: the metric-affine $F(R)$, $F(T)$, $F(Q)$ and $F(R,\mathcal{D})$, $F(T,\mathcal{D})$, $F(Q,\mathcal{D})$ theories, the metric-affine $F(R,T)$ model (minimal MAMG-I), the metric-affine $F(R,Q)$ one (minimal MAMG-II), the metric-affine $F(T,Q)$ theory (minimal MAMG-III), and the metric-affine $F(R,T,\mathcal{D})$, $F(R,Q,\mathcal{D})$, $F(T,Q,\mathcal{D})$ models (corresponding, respectively, to MAMG-I, MAMG-II, and MAMG-III).
In each sub-case, if we remove the dependence of the matter Lagrangian on the general affine connection, we are left with the respective sub-case of the MG-V theory.

\section{MG-VI and MAMG-VI}\label{mamg6}

Here we start with the action of the MG-VI model, which reads \cite{Myrzakulov:2012ug}
\begin{equation}\label{MG6}
\mathcal{S}^{(\rm{VI})}[g,\Gamma,\varphi] = \frac{1}{2 \kappa} \int \sqrt{-g} d^4 x \left[ F(R,Q,\mathcal{T}) + 2 \kappa \mathcal{L}_{\text{m}} \right] \,,
\end{equation}
where $F=F(R,Q,\mathcal{T})$ is a generic function of the scalar curvature $R$ of the general affine connection $\Gamma$, $Q$ is the non-metricity scalar, and $\mathcal{T}$ is the energy-momentum trace.
The action \eqref{MG6} represents and estension of the $F(R,\mathcal{T})$, $F(Q)$, $F(R)$, and $F(Q,\mathcal{T})$ (cf. \cite{Xu:2019sbp}) theories. In particular, \eqref{MG6} is an extension of the action \eqref{MG2} of the MG-II model, as it includes a dependence on $\mathcal{T}$ in $F$. \\
The metric field equations of the theory are given by
\begin{equation}\label{deltagMG6}
- \frac{1}{2} g_{\mu \nu} F + F'_R R_{(\mu \nu)} + F'_Q L_{(\mu \nu)} + \hat{\nabla}_\lambda \left( F'_Q {J^\lambda}_{(\mu \nu)} \right) + g_{\mu \nu} \hat{\nabla}_\lambda \left( F'_Q \zeta^\lambda \right) + F'_{\mathcal{T}} \left( \Theta_{\mu \nu} + T_{\mu \nu} \right) = \kappa T_{\mu \nu} \,,
\end{equation}
where $\hat{\nabla}$ is given by \eqref{hatder}, $L_{\mu \nu}$, ${J^\lambda}_{\mu \nu}$, $\zeta^\lambda$ are defined in \eqref{nonmetquant}, while $\Theta_{\mu \nu}$ is defined in \eqref{thetamunu}. \\
The connection field equations read
\begin{equation}\label{deltagammaMG6}
{P_\lambda}^{\mu \nu} (F'_R) + F'_Q \left[ 2 {Q^{[\nu \mu]}}_\lambda - {Q_\lambda}^{\mu \nu} + \left( q^\nu - Q^\nu \right) \delta^\mu_\lambda + Q_\lambda g^{\mu \nu} + \frac{1}{2} Q^\mu \delta^\nu_\lambda \right] = 0 \,,
\end{equation}
where ${P_\lambda}^{\mu \nu} (F'_R)$ is defined in \eqref{modpala}.
Note that \eqref{deltagammaMG6} coincides with the connection field equations \eqref{deltagammaMG2} of the MG-II model, which is a consequence of the fact that in the MG theories matter does not couple to the connection $\Gamma$.

\subsection{Metric-affine generalizations of the MG-VI model}

Now, we construct the metric-affine generalization of the MG-VI theory, that is we allow matter to couple to the general affine connection ${\Gamma^\lambda}_{\mu \nu}$.
On the same lines of what we have done in the previous sections, we write the MAMG-VI action as
\begin{equation}\label{aMAMG6}
\mathcal{S}^{(\rm{VI})}_{\text{MAMG}} = \frac{1}{2 \kappa} \int \sqrt{-g} d^4 x \left[ F(R,Q,\mathcal{T},\mathcal{D}) + 2 \kappa \mathcal{L}_{\text{m}} \right] \,,
\end{equation}
where, as usual, $\mathcal{D}$ is the divergence of the dilation current defined in \eqref{diverghyperm}. In \eqref{aMAMG6} the energy-momentum trace $\mathcal{T}$ and $\mathcal{D}$ are placed on an equal footing. \\
Varying the action \eqref{aMAMG6} with respect to the metric field we obtain
\begin{equation}\label{deltagMAMG6}
\begin{split}
& - \frac{1}{2} g_{\mu \nu} F + F'_R R_{(\mu \nu)} + F'_Q L_{(\mu \nu)} + \hat{\nabla}_\lambda \left( F'_Q {J^\lambda}_{(\mu \nu)} \right) + g_{\mu \nu} \hat{\nabla}_\lambda \left( F'_Q \zeta^\lambda \right) + F'_{\mathcal{T}} \left( \Theta_{\mu \nu} + T_{\mu \nu} \right) + F'_{\mathcal{D}} M_{\mu \nu} \\
& = \kappa T_{\mu \nu} \,,
\end{split}
\end{equation}
where we recall that $M_{\mu \nu}$ is given by \eqref{Mmunu}. \\
On the other hand, the variation of \eqref{aMAMG6} with respect to ${\Gamma^\lambda}_{\mu \nu}$ yields
\begin{equation}\label{deltagammaMAMG6}
\begin{split}
& {P_\lambda}^{\mu \nu} (F'_R) + F'_Q \left[ 2 {Q^{[\nu \mu]}}_\lambda - {Q_\lambda}^{\mu \nu} + \left( q^\nu - Q^\nu \right) \delta^\mu_\lambda + Q_\lambda g^{\mu \nu} + \frac{1}{2} Q^\mu \delta^\nu_\lambda \right] - {M_\lambda}^{\mu \nu \rho} \partial_\rho F'_{\mathcal{D}} \\
& = F'_{\mathcal{T}} {\Theta_\lambda}^{\mu \nu} + \kappa {\Delta_\lambda}^{\mu \nu} \,,
\end{split}
\end{equation}
where ${M_\lambda}^{\mu \nu \rho}$ and ${\Theta_\lambda}^{\mu \nu}$ are defined in \eqref{Mlmnr} and \eqref{Thetalmn}, respectively.
One can also consider a minimal metric-affine generalization of the MG-VI theory, excluding the $\mathcal{D}$ dependence in the function $F$. In this case, the metric field equations would coincide with \eqref{deltagMG6}, while the connection field equations would be
\begin{equation}\label{mindeltagammaMAMG6}
{P_\lambda}^{\mu \nu} (F'_R) + F'_Q \left[ 2 {Q^{[\nu \mu]}}_\lambda - {Q_\lambda}^{\mu \nu} + \left( q^\nu - Q^\nu \right) \delta^\mu_\lambda + Q_\lambda g^{\mu \nu} + \frac{1}{2} Q^\mu \delta^\nu_\lambda \right] = F'_{\mathcal{T}} {\Theta_\lambda}^{\mu \nu} + \kappa {\Delta_\lambda}^{\mu \nu} \,.
\end{equation}
The same result would follow by considering some specific matter fulfilling $\Delta^\nu=0$.
Let us also notice that if ${\Theta_\lambda}^{\mu \nu}=0$ the connection field equations \eqref{deltagammaMAMG6} reduce to those of the MAMG-II theory, namely \eqref{deltagammaMAMG2}. Analogously, if ${\Theta_\lambda}^{\mu \nu}=0$ then \eqref{mindeltagammaMAMG6} boil down to \eqref{mindeltagammaMAMG2}, that is to the connection field equations of the minimal MAMG-II model.

\subsection{Particular sub-cases of the MAMG-VI theory}

Regarding the MAMG-VI theory, besides the sub-case we have already mentioned above, corresponding to the minimal MAMG-VI model (minimal metric-affine generalization of MG-VI, that is the metric-affine $F(R,Q,\mathcal{T})$ theory), we also have the following sub-cases: the metric-affine $F(R)$, $F(Q)$ and $F(R,\mathcal{D})$, $F(Q,\mathcal{D})$ theories, the metric-affine $F(R,Q)$ model (minimal MAMG-II), the metric-affine $F(R,\mathcal{T})$ and $F(R,\mathcal{T},\mathcal{D})$ theories (cf. \eqref{frmtact} and \eqref{frmtdact}, respectively), and the metric-affine $F(R,Q,\mathcal{D})$ model (i.e., MAMG-II).
For the sake of completeness, let us finally report in the following other two sub-cases, namely the metric-affine $F(Q,\mathcal{T})$ and $F(Q,\mathcal{T},\mathcal{D})$ theories. In each of these sub-cases, removing the dependence of the matter Lagrangian on the general affine connection, one is left with the respective sub-cases of the MG-VI theory.

\subsubsection{Metric-affine $F(Q,\mathcal{T})$ theory}\label{fqmtgrav}

Restricting ourselves to $F(R,Q,\mathcal{T},\mathcal{D}) \to F(Q,\mathcal{T})$, the MAMG-VII model boils down to the metric-affine $F(Q,\mathcal{T})$ theory, whose action reads
\begin{equation}\label{fqmtact}
\mathcal{S}_{F(Q,\mathcal{T})} = \frac{1}{2 \kappa} \int \sqrt{-g} d^4 x \left[ F(Q,\mathcal{T}) + 2 \kappa \mathcal{L}_{\text{m}} \right] \,,
\end{equation}
where $Q$ is the non-metricity scalar and $\mathcal{T}$ the energy-momentum trace. The variation of the action \eqref{fqmtact} with respect to the metric and the affine connection gives the following set of field equations:
\begin{equation}
\begin{split}
& - \frac{1}{2} g_{\mu \nu} F + F'_Q L_{(\mu \nu)} + \hat{\nabla}_\lambda \left(F'_Q {J^\lambda}_{(\mu \nu)} \right) + g_{\mu \nu} \hat{\nabla}_\lambda \left(F'_Q \zeta^\lambda \right) + F'_{\mathcal{T}} \left(\Theta_{\mu \nu} + T_{\mu \nu} \right) = \kappa T_{\mu \nu} \,, \\
& F'_Q \left[ 2 {Q^{[\nu \mu]}}_\lambda - {Q_\lambda}^{\mu \nu} + \left( q^\nu - Q^\nu \right) \delta^\mu_\lambda + Q_\lambda g^{\mu \nu} + \frac{1}{2} Q^\mu \delta^\nu_\lambda \right] = F'_{\mathcal{T}} {\Theta_\lambda}^{\mu \nu} + \kappa {\Delta_\lambda}^{\mu \nu} \,,
\end{split}
\end{equation}
where $L_{\mu \nu}$, ${J^\lambda}_{\mu \nu}$, and $\zeta^\lambda$ have been defined in \eqref{nonmetquant}, while $\Theta_{\mu nu}$ and ${\Theta_\lambda}^{\mu \nu}$ are given by eqs. \eqref{thetamunu} and \eqref{Thetalmn}, respectively.

\subsubsection{Metric-affine $F(Q,\mathcal{T},\mathcal{D})$ theory}\label{fqmtdgrav}

On the other hand, considering $F(R,Q,\mathcal{T},\mathcal{D}) \to F(Q,\mathcal{T},\mathcal{D})$, namely excluding the dependence on the scalar curvature $R$ in the MAMG-VI model, the latter reduces to the metric-affine $F(Q,\mathcal{T},\mathcal{D})$ theory, which, in turn, is an extension of the $F(Q,\mathcal{T})$ theory above involving a dependence on the divergence of the dilation current in $F$ as well. The action of the metric-affine $F(Q,\mathcal{T},\mathcal{D})$ model has the following form:
\begin{equation}\label{fqmtdact}
\mathcal{S}_{F(Q,\mathcal{T},\mathcal{D})} = \frac{1}{2 \kappa} \int \sqrt{-g} d^4 x \left[ F(Q,\mathcal{T},\mathcal{D}) + 2 \kappa \mathcal{L}_{\text{m}} \right] \,,
\end{equation}
where $Q$ is the non-metricity scalar, $\mathcal{T}$ the energy-momentum trace, and $\mathcal{D}$ the divergence of the dilation current. Varying the action \eqref{fqmtdact} with respect to the metric and the general affine connection, we obtain the set of field equations
\begin{equation}
\begin{split}
& - \frac{1}{2} g_{\mu \nu} F + F'_Q L_{(\mu \nu)} + \hat{\nabla}_\lambda \left(F'_Q {J^\lambda}_{(\mu \nu)} \right) + g_{\mu \nu} \hat{\nabla}_\lambda \left(F'_Q \zeta^\lambda \right) + F'_{\mathcal{T}} \left(\Theta_{\mu \nu} + T_{\mu \nu} \right) + F'_{\mathcal{D}} M_{\mu \nu} = \kappa T_{\mu \nu} \,, \\
& F'_Q \left[ 2 {Q^{[\nu \mu]}}_\lambda - {Q_\lambda}^{\mu \nu} + \left( q^\nu - Q^\nu \right) \delta^\mu_\lambda + Q_\lambda g^{\mu \nu} + \frac{1}{2} Q^\mu \delta^\nu_\lambda \right] - {M_\lambda}^{\mu \nu \rho} \partial_\rho F'_{\mathcal{D}} = F'_{\mathcal{T}} {\Theta_\lambda}^{\mu \nu} + \kappa {\Delta_\lambda}^{\mu \nu} \,,
\end{split}
\end{equation}
with $M_{\mu \nu}$ and ${M_\lambda}^{\mu \nu \rho}$ given by \eqref{Mmunu} and \eqref{Mlmnr}, respectively.

\section{MG-VII and MAMG-VII}\label{mamg7}

In this section we first consider the MG-VII theory and subsequently generalize it by considering a metric-affine setup.
The MG-VII action \cite{Myrzakulov:2012ug} reads as follows:
\begin{equation}\label{MG7}
\mathcal{S}^{(\rm{VII})}[g,\Gamma,\varphi] = \frac{1}{2 \kappa} \int \sqrt{-g} d^4 x \left[ F(T,Q,\mathcal{T}) + 2 \kappa \mathcal{L}_{\text{m}} \right] \,,
\end{equation}
where $F=F(T,Q,\mathcal{T})$ is a generic function of the torsion and non-metricity scalars ($T$ and $Q$, respectively) and of the trace, $\mathcal{T}$, of the energy-momentum tensor. In fact, \eqref{MG7} is an extension of the action \eqref{MG3} of the MG-III model as it includes also a dependence on $\mathcal{T}$ in $F$. \\
By varying the action \eqref{MG7} with respect to the metric we get
\begin{equation}\label{deltagMG7}
\begin{split}
& - \frac{1}{2} g_{\mu \nu} F + F'_T \left( 2 S_{\nu \alpha \beta} {S_\mu}^{\alpha \beta} - S_{\alpha \beta \mu} {S^{\alpha \beta}}_\nu + 2 S_{\nu \alpha \beta} {S_\mu}^{\beta \alpha} - 4 S_\mu S_\nu \right) + F'_Q L_{(\mu \nu)} \\
& + \hat{\nabla}_\lambda \left( F'_Q {J^\lambda}_{(\mu \nu)} \right) + g_{\mu \nu} \hat{\nabla}_\lambda \left( F'_Q \zeta^\lambda \right) + F'_{\mathcal{T}} \left( \Theta_{\mu \nu} + T_{\mu \nu} \right)   = \kappa T_{\mu \nu} \,,
\end{split}
\end{equation}
where $\hat{\nabla}$ is given in \eqref{hatder}, $L_{\mu \nu}$, ${J^\lambda}_{\mu \nu}$ and $\zeta^\lambda$ are defined in \eqref{nonmetquant}, while $\Theta_{\mu \nu}$ is given by \eqref{thetamunu}. \\
The variation of \eqref{MG7} with respect to the general affine connection gives
\begin{equation}\label{deltagammaMG7}
2 F'_T \left({S^{\mu \nu}}_\lambda - 2 {S_\lambda}^{[\mu \nu]} - 4 S^{[\mu} \delta^{\nu]}_\lambda \right) + F'_Q \left[ 2 {Q^{[\nu \mu]}}_\lambda - {Q_\lambda}^{\mu \nu} + \left( q^\nu - Q^\nu \right) \delta^\mu_\lambda + Q_\lambda g^{\mu \nu} + \frac{1}{2} Q^\mu \delta^\nu_\lambda \right] = 0 \,.
\end{equation}
Let us observe that \eqref{deltagammaMG7} coincides with the connection field equations \eqref{deltagammaMG3} of the MG-III theory as in the MG models matter does not couple to the connection.
We can now move on to the metric-affine generalization of the MG-VII model.

\subsection{Metric-affine generalizations of the MG-VII model}

On the same lines of what we have done in the previous sections, let us write the MAMG-VII action as
\begin{equation}\label{aMAMG7}
\mathcal{S}^{(\rm{VII})}_{\text{MAMG}} = \frac{1}{2 \kappa} \int \sqrt{-g} d^4 x \left[ F(T,Q,\mathcal{T},\mathcal{D}) + 2 \kappa \mathcal{L}_{\text{m}} \right] \,,
\end{equation}
where $\mathcal{D}$ is given by \eqref{diverghyperm}. Again, $\mathcal{T}$ and $\mathcal{D}$ are placed on the same footing. \\
The metric field equations of the theory are
\begin{equation}\label{deltagMAMG7}
\begin{split}
& - \frac{1}{2} g_{\mu \nu} F + F'_T \left( 2 S_{\nu \alpha \beta} {S_\mu}^{\alpha \beta} - S_{\alpha \beta \mu} {S^{\alpha \beta}}_\nu + 2 S_{\nu \alpha \beta} {S_\mu}^{\beta \alpha} - 4 S_\mu S_\nu \right) + F'_Q L_{(\mu \nu)} \\
& + \hat{\nabla}_\lambda \left( F'_Q {J^\lambda}_{(\mu \nu)} \right) + g_{\mu \nu} \hat{\nabla}_\lambda \left( F'_Q \zeta^\lambda \right) + F'_{\mathcal{T}} \left( \Theta_{\mu \nu} + T_{\mu \nu} \right) + F'_{\mathcal{D}} M_{\mu \nu} = \kappa T_{\mu \nu} \,,
\end{split}
\end{equation}
where we recall that $M_{\mu \nu}$ is defined in \eqref{Mmunu}. \\
The connection field equations read as follows:
\begin{equation}\label{deltagammaMAMG7}
\begin{split}
& 2 F'_T \left({S^{\mu \nu}}_\lambda - 2 {S_\lambda}^{[\mu \nu]} - 4 S^{[\mu} \delta^{\nu]}_\lambda \right) + F'_Q \left[ 2 {Q^{[\nu \mu]}}_\lambda - {Q_\lambda}^{\mu \nu} + \left( q^\nu - Q^\nu \right) \delta^\mu_\lambda + Q_\lambda g^{\mu \nu} + \frac{1}{2} Q^\mu \delta^\nu_\lambda \right] \\
& - {M_\lambda}^{\mu \nu \rho} \partial_\rho F'_{\mathcal{D}} = F'_{\mathcal{T}} {\Theta_\lambda}^{\mu \nu} + \kappa {\Delta_\lambda}^{\mu \nu} \,,
\end{split}
\end{equation}
where ${M_\lambda}^{\mu \nu \rho}$ and ${\Theta_\lambda}^{\mu \nu}$ are respectively defined in eq. \eqref{Mlmnr} and eq. \eqref{Thetalmn}.
We can also consider a minimal metric-affine generalization of the MG-VII theory by excluding the $\mathcal{D}$ dependence in the function $F$. In this minimal case, the metric field equations coincide with \eqref{deltagMG7}, while the connection field equations are
\begin{equation}\label{mindeltagammaMAMG7}
\begin{split}
& 2 F'_T \left({S^{\mu \nu}}_\lambda - 2 {S_\lambda}^{[\mu \nu]} - 4 S^{[\mu} \delta^{\nu]}_\lambda \right) + F'_Q \left[ 2 {Q^{[\nu \mu]}}_\lambda - {Q_\lambda}^{\mu \nu} + \left( q^\nu - Q^\nu \right) \delta^\mu_\lambda + Q_\lambda g^{\mu \nu} + \frac{1}{2} Q^\mu \delta^\nu_\lambda \right] \\
& = F'_{\mathcal{T}} {\Theta_\lambda}^{\mu \nu} + \kappa {\Delta_\lambda}^{\mu \nu} \,.
\end{split}
\end{equation}
Analogously to what we have seen in the previous sections, the same result would follow by considering some specific matter fulfilling $\Delta^\nu=0$.
Observe that if ${\Theta_\lambda}^{\mu \nu}=0$ then the connection field equations \eqref{deltagammaMAMG7} reduce to \eqref{deltagammaMAMG3}, which are those of the MAMG-III theory. Similarly, if ${\Theta_\lambda}^{\mu \nu}=0$ then \eqref{mindeltagammaMAMG7} boil down to \eqref{mindeltagammaMAMG7}, namely to the connection field equations of the minimal MAMG-III theory.

We conclude this section by mentioning that, regarding MAMG-VII, besides the sub-case we have already mentioned above, corresponding to the minimal MAMG-VII model (minimal metric-affine generalization of MG-VII, that is the metric-affine $F(T,Q,\mathcal{T})$ theory), we also have the following sub-cases: the metric-affine $F(T)$, $F(Q)$ and $F(T,\mathcal{D})$, $F(Q,\mathcal{D})$ theories, the metric-affine $F(T,Q)$ model (minimal MAMG-III), the metric-affine $F(T,\mathcal{T})$ and $F(T,\mathcal{T},\mathcal{D})$ theories (cf. \eqref{ftmtact} and \eqref{ftmtdact}, respectively), the metric-affine $F(Q,\mathcal{T})$ and $F(Q,\mathcal{T},\mathcal{D})$ models (see \eqref{fqmtact} and \eqref{fqmtdact}, respectively), and the metric-affine $F(T,Q,\mathcal{D})$ theory (i.e., MAMG-III).
In each sub-case, by removing the dependence of the matter Lagrangian on the general affine connection one is left with the respective sub-case of the MG-VII theory.

\section{MG-VIII and MAMG-VIII}\label{mamg8}

In this section we review the metric-affine generalization of the MG-VIII model, introduced in \cite{Iosifidis:2021kqo}. Before moving on to the MAG generalization, let us introduce the MG-VIII model, whose action is given by \cite{Myrzakulov:2012ug, Yesmakhanova:2021thj}
\begin{equation}\label{MG8}
\mathcal{S}^{(\rm{VIII})}[g,\Gamma,\varphi] = \frac{1}{2 \kappa} \int \sqrt{-g} d^4 x \left[ F(R,T,Q,\mathcal{T}) + 2 \kappa \mathcal{L}_{\text{m}} \right] \,,
\end{equation}
extending the $F(R)$, $F(T)$, $F(Q)$, $F(R,\mathcal{T})$ (cf. \cite{Harko:2011kv}), $F(T,\mathcal{T})$ (cf. \cite{Harko:2014aja}), and $F(Q, \mathcal{T})$ (cf. \cite{Xu:2019sbp}) theories.\footnote{In particular, assuming flatness (namely, ${R^\lambda}_{\mu \nu \rho}=0$) and vanishing non-metricity, \eqref{MG8} boils down to the torsionful theories of \cite{Myrzakulov:2010vz,Krssak:2015oua}, while demanding flatness and vanishing torsion one is left with the models of \cite{Nester:1998mp,BeltranJimenez:2018vdo}. On the other hand, imposing only teleparallelism in \eqref{MG8} we are left with the generalized theories of \cite{BeltranJimenez:2019odq,Jimenez:2021hai}.} The MG-VIII model is the most general of the MG theories, the function $F=F(R,T,Q,\mathcal{T})$ in \eqref{MG8} being a generic function of the scalar curvature $R$ of the general affine connection ${\Gamma^\lambda}_{\mu \nu}$, the torsion scalar $T$, the non-metricity scalar $Q$, and the energy-momentum trace $\mathcal{T}$. \\
The metric field equations of the theory read
\begin{equation}\label{deltagMG8}
\begin{split}
& - \frac{1}{2} g_{\mu \nu} F + F'_R R_{(\mu \nu)} + F'_T \left( 2 S_{\nu \alpha \beta} {S_\mu}^{\alpha \beta} - S_{\alpha \beta \mu} {S^{\alpha \beta}}_\nu + 2 S_{\nu \alpha \beta} {S_\mu}^{\beta \alpha} - 4 S_\mu S_\nu \right) + F'_Q L_{(\mu \nu)} \\
& + \hat{\nabla}_\lambda \left( F'_Q {J^\lambda}_{(\mu \nu)} \right) + g_{\mu \nu} \hat{\nabla}_\lambda \left( F'_Q \zeta^\lambda \right) + F'_{\mathcal{T}} \left( \Theta_{\mu \nu} + T_{\mu \nu} \right) = \kappa T_{\mu \nu} \,,
\end{split}
\end{equation}
where $\hat{\nabla}$ is defined in \eqref{hatder}, while $L_{\mu \nu}$ and the densities ${J^\lambda}_{\mu \nu}$ and $\zeta^\lambda$ have been defined in \eqref{nonmetquant}, and $\Theta_{\mu \nu}$ is given by \eqref{thetamunu}. \\
On the other hand, the connection field equations are
\begin{equation}\label{deltagammaMG8}
\begin{split}
& {P_\lambda}^{\mu \nu} (F'_R) + 2 F'_T \left({S^{\mu \nu}}_\lambda - 2 {S_\lambda}^{[\mu \nu]} - 4 S^{[\mu} \delta^{\nu]}_\lambda \right) \\
& + F'_Q \left[ 2 {Q^{[\nu \mu]}}_\lambda - {Q_\lambda}^{\mu \nu} + \left( q^\nu - Q^\nu \right) \delta^\mu_\lambda + Q_\lambda g^{\mu \nu} + \frac{1}{2} Q^\mu \delta^\nu_\lambda \right] = 0 \,,
\end{split}
\end{equation}
where ${P_\lambda}^{\mu \nu} (F'_R)$ is the modified Palatini tensor defined in \eqref{modpala}. Eq. \eqref{deltagammaMG8} coincides with \eqref{deltagammaMG5} obtained for the MG-V model, due to the fact that in the MG theories matter does not couple to the connection.

\subsection{Metric-affine generalizations of the MG-VIII model}

The MAMG-VIII action, generalizing to the metric-affine case \eqref{MG8}, is \cite{Iosifidis:2021kqo}
\begin{equation}\label{aMAMG8}
\mathcal{S}^{(\rm{VIII})}_{\text{MAMG}} = \frac{1}{2 \kappa} \int \sqrt{-g} d^4 x \left[ F(R,T,Q,\mathcal{T},\mathcal{D}) + 2 \kappa \mathcal{L}_{\text{m}} \right] \,,
\end{equation}
where the divergence of the dilation current $\mathcal{D}$ is given by \eqref{diverghyperm} and in \eqref{aMAMG8} it is placed on the same footing of $\mathcal{T}$. \\
The variation of the action \eqref{aMAMG8} with respect to the metric yields
\begin{equation}\label{deltagMAMG8}
\begin{split}
& - \frac{1}{2} g_{\mu \nu} F + F'_R R_{(\mu \nu)} + F'_T \left( 2 S_{\nu \alpha \beta} {S_\mu}^{\alpha \beta} - S_{\alpha \beta \mu} {S^{\alpha \beta}}_\nu + 2 S_{\nu \alpha \beta} {S_\mu}^{\beta \alpha} - 4 S_\mu S_\nu \right) + F'_Q L_{(\mu \nu)} \\
& + \hat{\nabla}_\lambda \left( F'_Q {J^\lambda}_{(\mu \nu)} \right) + g_{\mu \nu} \hat{\nabla}_\lambda \left( F'_Q \zeta^\lambda \right) + F'_{\mathcal{T}} \left( \Theta_{\mu \nu} + T_{\mu \nu} \right) + F'_{\mathcal{D}} M_{\mu \nu} = \kappa T_{\mu \nu} \,,
\end{split}
\end{equation}
where, in particular, $M_{\mu \nu}$ is given by \eqref{Mmunu}. \\
On the other hand, from the variation of the action \eqref{aMAMG8} with respect to the general affine connection ${\Gamma^\lambda}_{\mu \nu}$ we find
\begin{equation}\label{deltagammaMAMG8}
\begin{split}
& {P_\lambda}^{\mu \nu} (F'_R) + 2 F'_T \left({S^{\mu \nu}}_\lambda - 2 {S_\lambda}^{[\mu \nu]} - 4 S^{[\mu} \delta^{\nu]}_\lambda \right) \\
& + F'_Q \left[ 2 {Q^{[\nu \mu]}}_\lambda - {Q_\lambda}^{\mu \nu} + \left( q^\nu - Q^\nu \right) \delta^\mu_\lambda + Q_\lambda g^{\mu \nu} + \frac{1}{2} Q^\mu \delta^\nu_\lambda \right] - {M_\lambda}^{\mu \nu \rho} \partial_\rho F'_{\mathcal{D}} = F'_{\mathcal{T}} {\Theta_\lambda}^{\mu \nu} + \kappa {\Delta_\lambda}^{\mu \nu} \,,
\end{split}
\end{equation}
where ${M_\lambda}^{\mu \nu \rho}$ and ${\Theta_\lambda}^{\mu \nu}$ have been defined in \eqref{Mlmnr} and \eqref{Thetalmn}, respectively.
On the same lines of what we have seen for all the theories previously discussed, also here one might consider a minimal metric-affine generalization of the MG-VIII model by excluding the $\mathcal{D}$ dependence in the function $F$. The metric field equations in this minimal case coincide with \eqref{deltagMG8}, while the connection field equations read
\begin{equation}\label{mindeltagammaMAMG8}
\begin{split}
& {P_\lambda}^{\mu \nu} (F'_R) + 2 F'_T \left({S^{\mu \nu}}_\lambda - 2 {S_\lambda}^{[\mu \nu]} - 4 S^{[\mu} \delta^{\nu]}_\lambda \right) \\
& + F'_Q \left[ 2 {Q^{[\nu \mu]}}_\lambda - {Q_\lambda}^{\mu \nu} + \left( q^\nu - Q^\nu \right) \delta^\mu_\lambda + Q_\lambda g^{\mu \nu} + \frac{1}{2} Q^\mu \delta^\nu_\lambda \right] = F'_{\mathcal{T}} {\Theta_\lambda}^{\mu \nu} + \kappa {\Delta_\lambda}^{\mu \nu} \,.
\end{split}
\end{equation}
The same result would follow by considering some specific matter fulfilling $\Delta^\nu=0$.
Finally, observe that if ${\Theta_\lambda}^{\mu \nu}=0$ the connection field equations \eqref{deltagammaMAMG8} reduce to \eqref{deltagammaMAMG5}, namely to those of the MAMG-V theory. Analogously, if ${\Theta_\lambda}^{\mu \nu}=0$ eq. \eqref{mindeltagammaMAMG8} boil down to \eqref{mindeltagammaMAMG5}, that is to the connection field equations of the minimal MAMG-V model.

Let us conclude this section by mentioning that, concerning the MAMG-VIII theory, besides the sub-case we have already mentioned above, corresponding to minimal MAMG-VIII (minimal metric-affine generalization of MG-VIII, that is the metric-affine $F(R,T,Q,\mathcal{T})$ theory), we also have the following sub-cases: the metric-affine $F(R)$, $F(T)$, $F(Q)$, $F(R,\mathcal{D})$, $F(T,\mathcal{D})$, $F(Q,\mathcal{D})$, $F(R,\mathcal{T})$, $F(T,\mathcal{T})$, $F(Q,\mathcal{T})$, $F(R,\mathcal{T},\mathcal{D})$, $F(T,\mathcal{T},\mathcal{D})$, $F(Q,\mathcal{T},\mathcal{D})$ theories, the metric-affine $F(R,T)$, $F(R,Q)$, $F(T,Q)$ models (namely, the minimal MAMG-I, minimal MAMG-II, minimal MAMG-III theories, respectively), the metric-affine $F(R,T,\mathcal{D})$, $F(R,Q,\mathcal{D})$, $F(T,Q,\mathcal{D})$ models (MAMG-I, MAMG-II, MAMG-III, respectively), the metric-affine $F(R,T,\mathcal{T})$, $F(R,T,Q)$, $F(R,Q,\mathcal{T})$, $F(T,Q,\mathcal{T})$ theories (namely, minimal MAMG-IV, minimal MAMG-V, minimal MAMG-VI, minimal MAMG-VII, respectively), and the metric-affine $F(R,T,\mathcal{T},\mathcal{D})$, $F(R,T,Q,\mathcal{D})$, $F(R,Q,\mathcal{T},\mathcal{D})$, and $F(T,Q,\mathcal{T},\mathcal{D})$ models (i.e., respectively, MAMG-IV, MAMG-V, MAMG-VI, and MAMG-VII).

\section{Cosmological aspects of MAMG theories}\label{cosmolasp}

We can now move on to the of study cosmological aspects of the MAMG theories previously introduced. 
We focus on the case in which the function characterizing MAMG is linear and consider a homogeneous FLRW background in the presence of torsion and non-metricity (see Appendix \ref{appb} for a collection of useful formulas in this context). 
In particular, we derive the Friedmann equations for the aforementioned theories in this cosmological framework. \\
Let us start by recalling the second Friedmann equation (i.e., the acceleration equation, also known as Raychaudhuri equation) for general non-Riemannian cosmological setups, which was obtained in \cite{Iosifidis:2020zzp} and reads as follows:
\begin{equation}\label{secondF}
\frac{\ddot{a}}{a} = - \frac{1}{3} R_{\mu \nu} u^\mu u^\nu + 2 \left( \frac{\dot{a}}{a} \right) \Phi + 2 \dot{\Phi} + \left( \frac{\dot{a}}{a} \right) \left( A + \frac{C}{2} \right) + \frac{\dot{A}}{2} - \frac{A^2}{2} - \frac{1}{2} A C - 2 A \Phi - 2 C \Phi \,,
\end{equation}
where $a=a(t)$ is the scale factor of the universe, $u^\mu$ is the normalized four-velocity, and $\Phi=\Phi(t)$ and $A=A(t)$, $C=C(t)$ are functions appearing, respectively, in the expressions of torsion and non-metricity given by \eqref{tornonmetFLRW}, obtained, in fact, in the highly symmetric spacetime we are considering.
We now derive (variants of) the first Friedmann equation for the MAMG theories, focusing on the linear case.
Before moving on to the analysis of the various models, let us observe that, since $\sqrt{-g}\mathcal{D}$ is a total divergence, at the linear level the dilation current dependence in the function $F$ characterizing the theories does not contribute to the field equations. Thus, we will directly focus on the linear, minimal MAMG models, namely on those excluding the $\mathcal{D}$ dependence in $F$, as at the linear level the on-shell result would actually coincide with the one of the non-minimal cases.

\paragraph{Cosmology in linear MAMG-I \\} 

We consider the linear MAMG-I case in which
\begin{equation}\label{lin1}
F = R + \beta T \,,
\end{equation}
where $\beta$ is, in principle, a free parameter.\footnote{In fact, in this linear case one can write $F=\alpha R + \beta T$ and fix the normalization of the theory choosing $\alpha=1$.}
The metric field equations \eqref{deltagMAMG1} (actually, \eqref{deltagMG1}, as we have safely excluded the $\mathcal{D}$ dependence in $F$) now take the form
\begin{equation}\label{lindeltag1}
- \frac{1}{2} g_{\mu \nu} F + R_{(\mu \nu)} + \beta \left( 2 S_{\nu \alpha \beta} {S_\mu}^{\alpha \beta} - S_{\alpha \beta \mu} {S^{\alpha \beta}}_\nu + 2 S_{\nu \alpha \beta} {S_\mu}^{\beta \alpha} - 4 S_\mu S_\nu \right) = \kappa T_{\mu \nu} \,.
\end{equation}
Taking the trace of \eqref{lindeltag1},\footnote{Let us mention, here, that one might also contract \eqref{lindeltag1} with $u^\mu u^\nu$ and exploit the second Friedmann equation \eqref{secondF} in order to express everything in terms of the scale factor and the torsion and non-metricity parameters.} using the post-Riemannian expansion of the scalar curvature $R$ given in \eqref{prexpcosm} and the expressions of torsion and non-metricity in \eqref{tornonmetFLRW}, we get the following variant of the modified first Friedmann equation:
\begin{equation}\label{fried1}
\frac{\ddot{a}}{a} + \left(\frac{\dot{a}}{a} \right)^2 + \left( 1 + \beta \right) \left( 4 \Phi^2 - P^2 \right) + \frac{1}{8} \left[ 2 A^2 + B \left( C-A \right) \right] + \Phi \left( 2 A - B \right) + \dot{f}_1 + 3 H f_1 = - \frac{\kappa}{6} \mathcal{T} \,,
\end{equation}
where $\Phi=\Phi(t)$, $P=P(t)$ and $A=A(t)$, $B=B(t)$, $C=C(t)$ are the functions appearing in \eqref{tornonmetFLRW}, and we have also defined
\begin{equation}\label{f1}
f_1 := \frac{1}{2} \left( \frac{B}{2} - A - 4 \Phi \right) \,.
\end{equation}
In \cite{Iosifidis:2021kqo} it was provided a cosmological application of these results, considering the case in which the matter Lagrangian is the one for a scalar field $\phi$ coupled to torsion, restricting the theory to the torsionful case with vanishing non-metricity. We will review this application and further elaborate on it in Subsection \ref{applic}. \\
On the other hand, in this linear case the connection field equations of the model read
\begin{equation}\label{fieldeqscosm1}
{P_\lambda}^{\mu \nu} + 2 \beta \left({S^{\mu \nu}}_\lambda - 2 {S_\lambda}^{[\mu \nu]} - 4 S^{[\mu} \delta^{\nu]}_\lambda \right) = \kappa {\Delta_\lambda}^{\mu \nu} \,,
\end{equation}
where ${P_\lambda}^{\mu \nu}$ is the Palatini tensor defined in eq. \eqref{palatinidefin}. Let us also mention, here, that in the cosmological setup we are considering, upon use of eqs. \eqref{tornonmetFLRW} (first line) and \eqref{cosmPala}, eq. \eqref{fieldeqscosm1} takes the following form:
\begin{equation}
\begin{split}
& \left( \frac{1}{2} A + 4 \Phi - \frac{C}{2} \right) u_\lambda h^{\mu \nu} + \left( B - \frac{3}{2} A - 4 \Phi - \frac{C}{2} \right) u^\mu {h_\lambda}^\nu - \frac{B}{2} u^\nu {h^\mu}_\lambda - \frac{3}{2} B u_\lambda u^\mu u^\nu - 2 \varepsilon_{\lambda \phantom{\mu \nu} \rho}^{\phantom{\lambda} \mu \nu} u^\rho P \\
& + 2 \beta \left[ 4 \left( \delta_\lambda^{[\nu} {h^{\mu]}}_\rho u^\rho + 3 \delta_\lambda^{[\mu} u^{\nu]} + {h_{\lambda}}^{[\nu} u^{\mu]} \right) \Phi - \varepsilon_{\lambda \phantom{\mu \nu} \rho}^{\phantom{\lambda} \mu \nu} u^\rho P \right] = \kappa {\Delta_\lambda}^{\mu \nu} \,,
\end{split}
\end{equation}
where, in particular, the explicit expressions of ${\Delta_\lambda}^{\mu \nu}$ depend on the specific matter one might then consider (see, e.g., the explicit expression of the hypermomentum tensor in the presence of a cosmological hyperfluid given in Ref. \cite{Iosifidis:2020gth}).

\paragraph{Cosmology in linear MAMG-II \\} 

We now consider the linear MAMG-II case in which
\begin{equation}\label{lin2}
F = R + \gamma Q \,,
\end{equation}
where $\gamma$ is a free parameter.
Eq. \eqref{deltagMAMG2} now takes the following form:
\begin{equation}\label{lindeltag2}
- \frac{1}{2} g_{\mu \nu} F + R_{(\mu \nu)} + \gamma \left( L_{(\mu \nu)} + \hat{\nabla}_\lambda {J^\lambda}_{(\mu \nu)} + g_{\mu \nu} \hat{\nabla}_\lambda \zeta^\lambda \right) = \kappa T_{\mu \nu} \,.
\end{equation}
Taking the trace of \eqref{lindeltag2} and following the procedure described in the previous paragraph, one ends up with
\begin{equation}\label{fried2}
\frac{\ddot{a}}{a} + \left(\frac{\dot{a}}{a} \right)^2 + 4 \Phi^2 - P^2 + \frac{1}{8} \left[ 2 A^2 + B \left( C-A \right) \right] + \Phi \left( 2 A - B \right) + \dot{f}_2 + 3 H f_2 = - \frac{\kappa}{6} \mathcal{T} \,,
\end{equation}
where
\begin{equation}\label{f2}
f_2 := \frac{1}{2} \left[ \left( 1 - \gamma \right) \left( \frac{B}{2} - A \right) - 4 \Phi \right] \,.
\end{equation}
Eq. \eqref{fried2} is a variant of the modified first Friedmann equation for the linear MAMG-II theory. \\
Moreover, the connection field equations of the model now read
\begin{equation}
{P_\lambda}^{\mu \nu} + \gamma \left[ 2 {Q^{[\nu \mu]}}_\lambda - {Q_\lambda}^{\mu \nu} + \left( q^\nu - Q^\nu \right) \delta^\mu_\lambda + Q_\lambda g^{\mu \nu} + \frac{1}{2} Q^\mu \delta^\nu_\lambda \right] = \kappa {\Delta_\lambda}^{\mu \nu} \,.
\end{equation}
The latter, upon use of eqs. \eqref{tornonmetFLRW} (second line) and \eqref{cosmPala}, take the form
\begin{equation}
\begin{split}
& \left( \frac{1}{2} A + 4 \Phi - \frac{C}{2} \right) u_\lambda h^{\mu \nu} + \left( B - \frac{3}{2} A - 4 \Phi - \frac{C}{2} \right) u^\mu {h_\lambda}^\nu - \frac{B}{2} u^\nu {h^\mu}_\lambda - \frac{3}{2} B u_\lambda u^\mu u^\nu - 2 \varepsilon_{\lambda \phantom{\mu \nu} \rho}^{\phantom{\lambda} \mu \nu} u^\rho P \\
& + \gamma \Bigg [ \frac{B}{2} \delta_\lambda^\nu {h^\mu}_\rho u^\rho + B h_{\lambda \rho} h^{\mu \nu} u^\rho + \left( A - \frac{1}{2} B \right) \delta_\lambda^\mu {h^\nu}_\rho u^\rho + \left( 2 A - C \right) h^{\mu \nu} u_\lambda + \frac{1}{2} \left( 3 A - C \right) \delta_\lambda^\nu u^\mu \\
& - A {h_\lambda}^\nu u^\mu + 3 \left( \frac{1}{2} B - A \right) \delta_\lambda^\mu u^\nu + \left( A - B \right) {h_\lambda}^\mu u^\nu - B h_{\lambda \rho} u^\rho u^\mu u^\nu - 3 A u_\lambda u^\mu u^\nu \Bigg ] = \kappa {\Delta_\lambda}^{\mu \nu} 
\end{split}
\end{equation}
in the cosmological setup we are considering.

\paragraph{Cosmology in linear MAMG-III \\} 

Here we consider the linear MAMG-III case in which
\begin{equation}\label{lin3}
F = T + \gamma Q \,,
\end{equation}
where $\gamma$ is a free parameter.\footnote{In fact, in the linear case at hand one may write $F=\beta T + \gamma Q$ and fix the normalization of the theory choosing $\beta=1$.}
The metric field equations of the theory now read
\begin{equation}\label{lindeltag3}
\begin{split}
& - \frac{1}{2} g_{\mu \nu} F + 2 S_{\nu \alpha \beta} {S_\mu}^{\alpha \beta} - S_{\alpha \beta \mu} {S^{\alpha \beta}}_\nu + 2 S_{\nu \alpha \beta} {S_\mu}^{\beta \alpha} - 4 S_\mu S_\nu \\
& + \gamma \left( L_{(\mu \nu)} + \hat{\nabla}_\lambda {J^\lambda}_{(\mu \nu)} + g_{\mu \nu} \hat{\nabla}_\lambda \zeta^\lambda \right) = \kappa T_{\mu \nu} \,.
\end{split}
\end{equation}
Taking the trace of the latter and using the expressions of torsion and non-metricity in \eqref{tornonmetFLRW} we find
\begin{equation}\label{fried3}
4 \Phi^2 - P^2 + \dot{f}_3 + 3 H f_3 = - \frac{\kappa}{6} \mathcal{T} \,,
\end{equation}
where we have defined
\begin{equation}\label{f3}
f_3 := - \frac{\gamma}{2} \left( \frac{B}{2} - A \right) \,.
\end{equation}
Notice that the scale factor $a(t)$ appears in \eqref{fried3} only through the Hubble parameter $H:= \frac{\dot{a}}{a}$, due to the fact that the function $F$ in \eqref{lin3} does not depend on the scalar curvature $R$. Therefore, eq. \eqref{fried3} gives the expression of $H$ in terms of the other parameters of the linear MAMG-III theory and of $\mathcal{T}$. In particular, this has to be so since we are not in the realm of a gauge theory of gravity. In the case of a gauge theory of gravity, in fact, in the tetrads formalism one can fix the form of the affine connection from the very beginning in terms of the dynamical vielbein and find the consequent effective Friedmann equations. \\
On the other hand, let us also mention that the connection field equations of the model now become
\begin{equation}\label{fieldeqscosm3}
\begin{split}
& 2 \beta \left({S^{\mu \nu}}_\lambda - 2 {S_\lambda}^{[\mu \nu]} - 4 S^{[\mu} \delta^{\nu]}_\lambda \right) + \gamma \left[ 2 {Q^{[\nu \mu]}}_\lambda - {Q_\lambda}^{\mu \nu} + \left( q^\nu - Q^\nu \right) \delta^\mu_\lambda + Q_\lambda g^{\mu \nu} + \frac{1}{2} Q^\mu \delta^\nu_\lambda \right] \\
& = \kappa {\Delta_\lambda}^{\mu \nu} \,.
\end{split}
\end{equation}
In the cosmological setup we are considering, eq. \eqref{fieldeqscosm3} takes the form
\begin{equation}
\begin{split}
& 2 \beta \left[ 4 \left( \delta_\lambda^{[\nu} {h^{\mu]}}_\rho u^\rho + 3 \delta_\lambda^{[\mu} u^{\nu]} + {h_{\lambda}}^{[\nu} u^{\mu]} \right) \Phi - \varepsilon_{\lambda \phantom{\mu \nu} \rho}^{\phantom{\lambda} \mu \nu} u^\rho P \right] + \gamma \Bigg [ \frac{B}{2} \delta_\lambda^\nu {h^\mu}_\rho u^\rho + B h_{\lambda \rho} h^{\mu \nu} u^\rho \\
& + \left( A - \frac{1}{2} B \right) \delta_\lambda^\mu {h^\nu}_\rho u^\rho + \left( 2 A - C \right) h^{\mu \nu} u_\lambda + \frac{1}{2} \left( 3 A - C \right) \delta_\lambda^\nu u^\mu - A {h_\lambda}^\nu u^\mu + 3 \left( \frac{1}{2} B - A \right) \delta_\lambda^\mu u^\nu \\
& + \left( A - B \right) {h_\lambda}^\mu u^\nu - B h_{\lambda \rho} u^\rho u^\mu u^\nu - 3 A u_\lambda u^\mu u^\nu \Bigg ] = \kappa {\Delta_\lambda}^{\mu \nu} \,,
\end{split}
\end{equation}
upon use of eqs. \eqref{tornonmetFLRW} and \eqref{cosmPala}.

\paragraph{Cosmology in linear MAMG-IV \\} 

Let us now move on to the linear MAMG-IV case in which
\begin{equation}\label{lin4}
F = R + \beta T + \mu \mathcal{T} \,,
\end{equation}
where $\beta$ and $\mu$ are free parameters.
The metric field equations of the theory take the following form:
\begin{equation}\label{lindeltag4}
- \frac{1}{2} g_{\mu \nu} F + R_{(\mu \nu)} + \beta \left( 2 S_{\nu \alpha \beta} {S_\mu}^{\alpha \beta} - S_{\alpha \beta \mu} {S^{\alpha \beta}}_\nu + 2 S_{\nu \alpha \beta} {S_\mu}^{\beta \alpha} - 4 S_\mu S_\nu \right) + \mu \left( \Theta_{\mu \nu} + T_{\mu \nu} \right) = \kappa T_{\mu \nu} \,.
\end{equation}
Taking the trace of \eqref{lindeltag4}, using the post-Riemannian expansion of $R$, given by eq. \eqref{prexpcosm}, along with \eqref{tornonmetFLRW}, we obtain a variant of the modified first Friedmann equation of the linear MAMG-IV theory, that is
\begin{equation}\label{fried4}
\begin{split}
& \frac{\ddot{a}}{a} + \left(\frac{\dot{a}}{a} \right)^2 + \left( 1 + \beta \right) \left( 4 \Phi^2 - P^2 \right) + \frac{1}{8} \left[ 2 A^2 + B \left( C-A \right) \right] + \Phi \left( 2 A - B \right) + \dot{f}_1 + 3 H f_1 \\
& = \frac{\mu}{6} \left( \Theta - \mathcal{T} \right) - \frac{\kappa}{6} \mathcal{T} \,,
\end{split}
\end{equation}
where $f_1$ is defined in \eqref{f1} and
\begin{equation}\label{thetatr}
\Theta := \Theta_{\mu \nu} g^{\mu \nu} \,.
\end{equation}
The first term on the right-hand side of \eqref{fried4} is consequence of the (linear) $\mathcal{T}$ dependence in \eqref{lin4}, while all the other terms in \eqref{fried4} coincide with those appearing in eq. \eqref{fried1}. \\
Furthermore, in this linear case the connection field equations of the theory read as follows:
\begin{equation}
{P_\lambda}^{\mu \nu} + 2 \beta \left({S^{\mu \nu}}_\lambda - 2 {S_\lambda}^{[\mu \nu]} - 4 S^{[\mu} \delta^{\nu]}_\lambda \right) = \mu {\Theta_\lambda}^{\mu \nu} + \kappa {\Delta_\lambda}^{\mu \nu} \,.
\end{equation}
The latter, in the cosmological scenario we are considering, upon use of eqs. \eqref{tornonmetFLRW} (first line) and \eqref{cosmPala}, become
\begin{equation}
\begin{split}
& \left( \frac{1}{2} A + 4 \Phi - \frac{C}{2} \right) u_\lambda h^{\mu \nu} + \left( B - \frac{3}{2} A - 4 \Phi - \frac{C}{2} \right) u^\mu {h_\lambda}^\nu - \frac{B}{2} u^\nu {h^\mu}_\lambda - \frac{3}{2} B u_\lambda u^\mu u^\nu - 2 \varepsilon_{\lambda \phantom{\mu \nu} \rho}^{\phantom{\lambda} \mu \nu} u^\rho P \\
& + 2 \beta \left[ 4 \left( \delta_\lambda^{[\nu} {h^{\mu]}}_\rho u^\rho + 3 \delta_\lambda^{[\mu} u^{\nu]} + {h_{\lambda}}^{[\nu} u^{\mu]} \right) \Phi - \varepsilon_{\lambda \phantom{\mu \nu} \rho}^{\phantom{\lambda} \mu \nu} u^\rho P \right] = \mu {\Theta_\lambda}^{\mu \nu} + \kappa {\Delta_\lambda}^{\mu \nu} \,,
\end{split}
\end{equation}
where, in particular, the explicit expressions of ${\Theta_\lambda}^{\mu \nu}$ and ${\Delta_\lambda}^{\mu \nu}$ depend on the specific matter one might then consider.

\paragraph{Cosmology in linear MAMG-V \\} 

In this paragraph we consider the linear MAMG-V case in which
\begin{equation}\label{lin5}
F = R + \beta T + \gamma Q \,,
\end{equation}
where $\beta$ and $\gamma$ are free parameters.
The metric field equations of the theory read as follows:
\begin{equation}\label{lindeltag5}
\begin{split}
& - \frac{1}{2} g_{\mu \nu} F + R_{(\mu \nu)} + \beta \left( 2 S_{\nu \alpha \beta} {S_\mu}^{\alpha \beta} - S_{\alpha \beta \mu} {S^{\alpha \beta}}_\nu + 2 S_{\nu \alpha \beta} {S_\mu}^{\beta \alpha} - 4 S_\mu S_\nu \right) \\
& + \gamma \left( L_{(\mu \nu)} + \hat{\nabla}_\lambda {J^\lambda}_{(\mu \nu)} + g_{\mu \nu} \hat{\nabla}_\lambda \zeta^\lambda \right) = \kappa T_{\mu \nu} \,.
\end{split}
\end{equation}
Taking the trace of the latter and following what we have done in the previous paragraph, we are left with a variant of the modified first Friedmann equation of the linear MAMG-V theory, that is
\begin{equation}\label{fried5}
\frac{\ddot{a}}{a} + \left(\frac{\dot{a}}{a} \right)^2 + \left( 1 + \beta \right) \left( 4 \Phi^2 - P^2 \right) + \frac{1}{8} \left[ 2 A^2 + B \left( C-A \right) \right] + \Phi \left( 2 A - B \right) + \dot{f}_2 + 3 H f_2 = - \frac{\kappa}{6} \mathcal{T} \,,
\end{equation}
where $f_2$ is defined in \eqref{f2}. \\
Besides, in the linear case \eqref{lin5} the connection field equations of the theory become
\begin{equation}
\begin{split}
& {P_\lambda}^{\mu \nu} + 2 \beta \left({S^{\mu \nu}}_\lambda - 2 {S_\lambda}^{[\mu \nu]} - 4 S^{[\mu} \delta^{\nu]}_\lambda \right) \\
& + \gamma \left[ 2 {Q^{[\nu \mu]}}_\lambda - {Q_\lambda}^{\mu \nu} + \left( q^\nu - Q^\nu \right) \delta^\mu_\lambda + Q_\lambda g^{\mu \nu} + \frac{1}{2} Q^\mu \delta^\nu_\lambda \right] = \kappa {\Delta_\lambda}^{\mu \nu} \,.
\end{split}
\end{equation}
The latter, in the cosmological setup we are considering, take the form
\begin{equation}
\begin{split}
& \left( \frac{1}{2} A + 4 \Phi - \frac{C}{2} \right) u_\lambda h^{\mu \nu} + \left( B - \frac{3}{2} A - 4 \Phi - \frac{C}{2} \right) u^\mu {h_\lambda}^\nu - \frac{B}{2} u^\nu {h^\mu}_\lambda - \frac{3}{2} B u_\lambda u^\mu u^\nu - 2 \varepsilon_{\lambda \phantom{\mu \nu} \rho}^{\phantom{\lambda} \mu \nu} u^\rho P \\
& + 2 \beta \left[ 4 \left( \delta_\lambda^{[\nu} {h^{\mu]}}_\rho u^\rho + 3 \delta_\lambda^{[\mu} u^{\nu]} + {h_{\lambda}}^{[\nu} u^{\mu]} \right) \Phi - \varepsilon_{\lambda \phantom{\mu \nu} \rho}^{\phantom{\lambda} \mu \nu} u^\rho P \right] + \gamma \Bigg [ \frac{B}{2} \delta_\lambda^\nu {h^\mu}_\rho u^\rho + B h_{\lambda \rho} h^{\mu \nu} u^\rho \\
& + \left( A - \frac{1}{2} B \right) \delta_\lambda^\mu {h^\nu}_\rho u^\rho + \left( 2 A - C \right) h^{\mu \nu} u_\lambda + \frac{1}{2} \left( 3 A - C \right) \delta_\lambda^\nu u^\mu - A {h_\lambda}^\nu u^\mu + 3 \left( \frac{1}{2} B - A \right) \delta_\lambda^\mu u^\nu \\
& + \left( A - B \right) {h_\lambda}^\mu u^\nu - B h_{\lambda \rho} u^\rho u^\mu u^\nu - 3 A u_\lambda u^\mu u^\nu \Bigg ] = \kappa {\Delta_\lambda}^{\mu \nu}
\end{split}
\end{equation}
upon use of eqs. \eqref{tornonmetFLRW} and \eqref{cosmPala}.

\paragraph{Cosmology in linear MAMG-VI \\} 

We now consider the linear MAMG-VI theory in which
\begin{equation}\label{lin6}
F = R + \gamma Q + \mu \mathcal{T} \,,
\end{equation}
where $\gamma$ and $\mu$ are free parameters.
The metric field equations of the model at hand read
\begin{equation}\label{lindeltag6}
- \frac{1}{2} g_{\mu \nu} F + R_{(\mu \nu)} + \gamma \left( L_{(\mu \nu)} + \hat{\nabla}_\lambda {J^\lambda}_{(\mu \nu)} + g_{\mu \nu} \hat{\nabla}_\lambda \zeta^\lambda \right) + \mu \left( \Theta_{\mu \nu} + T_{\mu \nu} \right) = \kappa T_{\mu \nu} \,.
\end{equation}
Taking the trace of \eqref{lindeltag6} and using \eqref{prexpcosm} and \eqref{tornonmetFLRW}, we arrive at
\begin{equation}\label{fried6}
\frac{\ddot{a}}{a} + \left(\frac{\dot{a}}{a} \right)^2 + 4 \Phi^2 - P^2 + \frac{1}{8} \left[ 2 A^2 + B \left( C-A \right) \right] + \Phi \left( 2 A - B \right) + \dot{f}_2 + 3 H f_2 = \frac{\mu}{6} \left( \Theta - \mathcal{T} \right) - \frac{\kappa}{6} \mathcal{T} \,,
\end{equation}
where $f_2$ and $\Theta$ are defined, respectively, in \eqref{f2} and \eqref{thetatr}.
Eq. \eqref{fried6} is a variant of the modified first Friedmann equation of the linear MAMG-VI model. \\
Besides, in the linear case at hand the connection field equations of the model read
\begin{equation}
{P_\lambda}^{\mu \nu} + \gamma \left[ 2 {Q^{[\nu \mu]}}_\lambda - {Q_\lambda}^{\mu \nu} + \left( q^\nu - Q^\nu \right) \delta^\mu_\lambda + Q_\lambda g^{\mu \nu} + \frac{1}{2} Q^\mu \delta^\nu_\lambda \right] = \mu {\Theta_\lambda}^{\mu \nu} + \kappa {\Delta_\lambda}^{\mu \nu} \,,
\end{equation}
which, in the cosmological setup we are considering, take the form
\begin{equation}
\begin{split}
& \left( \frac{1}{2} A + 4 \Phi - \frac{C}{2} \right) u_\lambda h^{\mu \nu} + \left( B - \frac{3}{2} A - 4 \Phi - \frac{C}{2} \right) u^\mu {h_\lambda}^\nu - \frac{B}{2} u^\nu {h^\mu}_\lambda - \frac{3}{2} B u_\lambda u^\mu u^\nu - 2 \varepsilon_{\lambda \phantom{\mu \nu} \rho}^{\phantom{\lambda} \mu \nu} u^\rho P \\
& + \gamma \Bigg [ \frac{B}{2} \delta_\lambda^\nu {h^\mu}_\rho u^\rho + B h_{\lambda \rho} h^{\mu \nu} u^\rho + \left( A - \frac{1}{2} B \right) \delta_\lambda^\mu {h^\nu}_\rho u^\rho + \left( 2 A - C \right) h^{\mu \nu} u_\lambda + \frac{1}{2} \left( 3 A - C \right) \delta_\lambda^\nu u^\mu \\
& - A {h_\lambda}^\nu u^\mu + 3 \left( \frac{1}{2} B - A \right) \delta_\lambda^\mu u^\nu + \left( A - B \right) {h_\lambda}^\mu u^\nu - B h_{\lambda \rho} u^\rho u^\mu u^\nu - 3 A u_\lambda u^\mu u^\nu \Bigg ] \\
& = \mu {\Theta_\lambda}^{\mu \nu} + \kappa {\Delta_\lambda}^{\mu \nu} \,,
\end{split}
\end{equation}
where we have used eqs. \eqref{tornonmetFLRW} (second line) and \eqref{cosmPala}.

\paragraph{Cosmology in linear MAMG-VII \\} 

Here we focus on the linear MAMG-VII case in which
\begin{equation}\label{lin7}
F = T + \gamma Q + \mu \mathcal{T} \,,
\end{equation}
where $\gamma$ and $\mu$ are a free parameters.
The metric field equations of the theory become
\begin{equation}\label{lindeltag7}
\begin{split}
& - \frac{1}{2} g_{\mu \nu} F + 2 S_{\nu \alpha \beta} {S_\mu}^{\alpha \beta} - S_{\alpha \beta \mu} {S^{\alpha \beta}}_\nu + 2 S_{\nu \alpha \beta} {S_\mu}^{\beta \alpha} - 4 S_\mu S_\nu \\
& + \gamma \left( L_{(\mu \nu)} + \hat{\nabla}_\lambda {J^\lambda}_{(\mu \nu)} + g_{\mu \nu} \hat{\nabla}_\lambda \zeta^\lambda \right) + \mu \left( \Theta_{\mu \nu} + T_{\mu \nu} \right) = \kappa T_{\mu \nu} \,.
\end{split}
\end{equation}
Taking the trace of eq. \eqref{lindeltag7} and using the expressions of torsion and non-metricity in \eqref{tornonmetFLRW}, we obtain
\begin{equation}\label{fried7}
4 \Phi^2 - P^2 + \dot{f}_3 + 3 H f_3 = \frac{\mu}{6} \left( \Theta - \mathcal{T} \right) - \frac{\kappa}{6} \mathcal{T} \,,
\end{equation}
where $f_3$ is given by \eqref{f3}, while $\Theta$ is defined in \eqref{thetatr}.
Observe that in this case, analogously to what happens in the linear MAMG-III theory, the scale factor $a(t)$ appears in \eqref{fried7} only through the Hubble parameter $H$, which is due to the fact that the function $F$ in \eqref{lin7} does not depend on the scalar curvature $R$ of the general affine connection $\Gamma$. One could then apply arguments analogous to those made in the case of the MAMG-III cosmology. \\
Moreover, in the linear case given by eq. \eqref{lin7}, the connection field equations of the theory take the following form:
\begin{equation}
\begin{split}
& 2 \beta \left({S^{\mu \nu}}_\lambda - 2 {S_\lambda}^{[\mu \nu]} - 4 S^{[\mu} \delta^{\nu]}_\lambda \right) + \gamma \left[ 2 {Q^{[\nu \mu]}}_\lambda - {Q_\lambda}^{\mu \nu} + \left( q^\nu - Q^\nu \right) \delta^\mu_\lambda + Q_\lambda g^{\mu \nu} + \frac{1}{2} Q^\mu \delta^\nu_\lambda \right] \\
& = \mu {\Theta_\lambda}^{\mu \nu} + \kappa {\Delta_\lambda}^{\mu \nu} \,,
\end{split}
\end{equation}
which, exploiting once again eqs. \eqref{tornonmetFLRW} and \eqref{cosmPala}, become
\begin{equation}
\begin{split}
& + 2 \beta \left[ 4 \left( \delta_\lambda^{[\nu} {h^{\mu]}}_\rho u^\rho + 3 \delta_\lambda^{[\mu} u^{\nu]} + {h_{\lambda}}^{[\nu} u^{\mu]} \right) \Phi - \varepsilon_{\lambda \phantom{\mu \nu} \rho}^{\phantom{\lambda} \mu \nu} u^\rho P \right] + \gamma \Bigg [ \frac{B}{2} \delta_\lambda^\nu {h^\mu}_\rho u^\rho + B h_{\lambda \rho} h^{\mu \nu} u^\rho \\
& + \left( A - \frac{1}{2} B \right) \delta_\lambda^\mu {h^\nu}_\rho u^\rho + \left( 2 A - C \right) h^{\mu \nu} u_\lambda + \frac{1}{2} \left( 3 A - C \right) \delta_\lambda^\nu u^\mu - A {h_\lambda}^\nu u^\mu + 3 \left( \frac{1}{2} B - A \right) \delta_\lambda^\mu u^\nu \\
& + \left( A - B \right) {h_\lambda}^\mu u^\nu - B h_{\lambda \rho} u^\rho u^\mu u^\nu - 3 A u_\lambda u^\mu u^\nu \Bigg ] = \mu {\Theta_\lambda}^{\mu \nu} + \kappa {\Delta_\lambda}^{\mu \nu}
\end{split}
\end{equation}
in the cosmological scenario we are considering.

\paragraph{Cosmology in linear MAMG-VIII \\} 

Finally, let us report in the following the cosmological aspects of the linear MAMG-VIII theory. In this case, we have
\begin{equation}\label{lin8}
F = R + \beta T + \gamma Q + \mu \mathcal{T} \,,
\end{equation} 
$\beta$, $\gamma$, and $\mu$ being free parameters, and the metric field equations of the theory take the form
\begin{equation}\label{lindeltag8}
\begin{split}
& - \frac{1}{2} g_{\mu \nu} F + R_{(\mu \nu)} + \beta \left( 2 S_{\nu \alpha \beta} {S_\mu}^{\alpha \beta} - S_{\alpha \beta \mu} {S^{\alpha \beta}}_\nu + 2 S_{\nu \alpha \beta} {S_\mu}^{\beta \alpha} - 4 S_\mu S_\nu \right) \\
& + \gamma \left( L_{(\mu \nu)} + \hat{\nabla}_\lambda {J^\lambda}_{(\mu \nu)} + g_{\mu \nu} \hat{\nabla}_\lambda \zeta^\lambda \right) + \mu \left( \Theta_{\mu \nu} + T_{\mu \nu} \right) = \kappa T_{\mu \nu} \,.
\end{split}
\end{equation}
Taking the trace of \eqref{lindeltag8} and using \eqref{prexpcosm} and \eqref{tornonmetFLRW}, we arrive at
\begin{equation}\label{fried8}
\begin{split}
& \frac{\ddot{a}}{a} + \left(\frac{\dot{a}}{a} \right)^2 + \left( 1 + \beta \right) \left( 4 \Phi^2 - P^2 \right) + \frac{1}{8} \left[ 2 A^2 + B \left( C-A \right) \right] + \Phi \left( 2 A - B \right) + \dot{f}_2 + 3 H f_2 \\
& = \frac{\mu}{6} \left( \Theta - \mathcal{T} \right) - \frac{\kappa}{6} \mathcal{T} \,,
\end{split}
\end{equation}
where $f_2$ and $\Theta$ are defined in \eqref{f2} and \eqref{thetatr}, respectively.
Eq. \eqref{fried8} is a variant of the modified first Friedmann equation which, together with the acceleration equation \eqref{secondF}, rules the cosmology of the linear MAMG-VIII model. \\
Besides, in this linear case the connection field equations of the theory read
\begin{equation}
\begin{split}
& {P_\lambda}^{\mu \nu} + 2 \beta \left({S^{\mu \nu}}_\lambda - 2 {S_\lambda}^{[\mu \nu]} - 4 S^{[\mu} \delta^{\nu]}_\lambda \right) \\
& + \gamma \left[ 2 {Q^{[\nu \mu]}}_\lambda - {Q_\lambda}^{\mu \nu} + \left( q^\nu - Q^\nu \right) \delta^\mu_\lambda + Q_\lambda g^{\mu \nu} + \frac{1}{2} Q^\mu \delta^\nu_\lambda \right] = \mu {\Theta_\lambda}^{\mu \nu} + \kappa {\Delta_\lambda}^{\mu \nu} \,.
\end{split}
\end{equation}
The latter, in the cosmological setup we are considering, take the form
\begin{equation}
\begin{split}
& \left( \frac{1}{2} A + 4 \Phi - \frac{C}{2} \right) u_\lambda h^{\mu \nu} + \left( B - \frac{3}{2} A - 4 \Phi - \frac{C}{2} \right) u^\mu {h_\lambda}^\nu - \frac{B}{2} u^\nu {h^\mu}_\lambda - \frac{3}{2} B u_\lambda u^\mu u^\nu - 2 \varepsilon_{\lambda \phantom{\mu \nu} \rho}^{\phantom{\lambda} \mu \nu} u^\rho P \\
& + 2 \beta \left[ 4 \left( \delta_\lambda^{[\nu} {h^{\mu]}}_\rho u^\rho + 3 \delta_\lambda^{[\mu} u^{\nu]} + {h_{\lambda}}^{[\nu} u^{\mu]} \right) \Phi - \varepsilon_{\lambda \phantom{\mu \nu} \rho}^{\phantom{\lambda} \mu \nu} u^\rho P \right] + \gamma \Bigg [ \frac{B}{2} \delta_\lambda^\nu {h^\mu}_\rho u^\rho + B h_{\lambda \rho} h^{\mu \nu} u^\rho \\
& + \left( A - \frac{1}{2} B \right) \delta_\lambda^\mu {h^\nu}_\rho u^\rho + \left( 2 A - C \right) h^{\mu \nu} u_\lambda + \frac{1}{2} \left( 3 A - C \right) \delta_\lambda^\nu u^\mu - A {h_\lambda}^\nu u^\mu + 3 \left( \frac{1}{2} B - A \right) \delta_\lambda^\mu u^\nu \\
& + \left( A - B \right) {h_\lambda}^\mu u^\nu - B h_{\lambda \rho} u^\rho u^\mu u^\nu - 3 A u_\lambda u^\mu u^\nu \Bigg ] = \mu {\Theta_\lambda}^{\mu \nu} + \kappa {\Delta_\lambda}^{\mu \nu} \,,
\end{split}
\end{equation}
where, as in the other models previously studied, we have used eqs. \eqref{tornonmetFLRW} and \eqref{cosmPala}.

\subsection{On the case of a scalar field coupled to torsion}\label{applic}

We shall now review and rediscuss the cosmological application in \cite{Iosifidis:2021kqo} to the case in which the matter Lagrangian is given by the one for a scalar field $\phi$ coupled to torsion (in particular, by means of the torsion vector $S^\mu$), where the non-metricity has been set to zero. The matter Lagrangian reads
\begin{equation}\label{scalLagr}
\mathcal{L}_{\text{m}} = - \frac{1}{2} g^{\mu \nu} \nabla_\mu \phi \nabla_\nu \phi - V (\phi) + \lambda_0 S^\mu \nabla_\mu \phi \,,
\end{equation}
where $\lambda_0$ is a constant parameter and we recall that $\nabla_\mu \phi = \partial_\mu \phi$. In particular, the authors of \cite{Iosifidis:2021kqo} considered the linear MAMG-I case \eqref{lin1}.
Hence, the full action of the theory is
\begin{equation}\label{scal1}
\mathcal{S} = \frac{1}{2 \kappa} \int \sqrt{-g} d^4 x \left[ R + \beta T + 2 \kappa \left( - \frac{1}{2} g^{\mu \nu} \nabla_\mu \phi \nabla_\nu \phi - V (\phi) + \lambda_0 S^\mu \nabla_\mu \phi \right) \right] \,.
\end{equation}
The variation of \eqref{scal1} with respect to the scalar field $\phi$ yields
\begin{equation}\label{eqscal}
\frac{1}{\sqrt{-g}} \partial_\mu \left[ \sqrt{-g} \left( \partial^\mu \phi - \lambda_0 S^\mu \right) \right] = \frac{\partial V}{\partial \phi} \,.
\end{equation}
Besides, the hypermomentum in this case reads\footnote{It can be computed by using eqs. \eqref{variationQSpart} (more precisely, the variation of the torsion vector with respect to the general affine connection).}
\begin{equation}\label{hypermscal}
{\Delta_\lambda}^{\mu \nu} = 2 \lambda_0 \delta^{[\mu}_\lambda \nabla^{\nu]} \phi \,.
\end{equation}
Varying the action \eqref{scal1} with respect to the general affine connection ${\Gamma^\lambda}_{\mu \nu}$ one finds
\begin{equation}\label{connfescal1}
{P_\lambda}^{\mu \nu} + 2 \beta \left({S^{\mu \nu}}_\lambda - 2 {S_\lambda}^{[\mu \nu]} - 4 S^{[\mu} \delta^{\nu]}_\lambda \right) = 2 \kappa \lambda_0 \delta^{[\mu}_\lambda \nabla^{\nu]} \phi \,,
\end{equation}
where ${P_\lambda}^{\mu \nu}$ is the Palatini tensor defined in eq. \eqref{palatinidefin}.
In particular, taking different contractions of \eqref{connfescal1}, we get
\begin{equation}\label{correctres}
\begin{split}
& S_\lambda = 0 \,, \\
& S^\mu = \frac{3 \kappa \lambda_0}{8 \beta} \partial^\mu \phi \,, \\
& (1+\beta) t_\lambda = 0 \,,
\end{split}
\end{equation}
where $t_\lambda$ is the torsion pseudo-vector (cf. Appendix \ref{appa}). The equations in \eqref{correctres} indicate that the torsion vector vanishes and, therefore, in order to have non-trivial dynamics for the scalar field, one should set $\lambda_0=0$, namely remove the coupling to torsion in \eqref{scalLagr}. Besides, the last of \eqref{correctres} implies that either $\beta=-1$ or $t_\lambda=0$. Fixing
\begin{equation}\label{choice}
\lambda_0 = 0 \,, \quad \beta = -1 \,,
\end{equation}
and plugging \eqref{correctres} back into \eqref{connfescal1}, the latter boils down to
\begin{equation}
{Z_{\lambda}}^{\mu \nu} = 0 \,,
\end{equation}
namely the traceless part of the torsion vanishes, while the only remaining torsion components are the the four of the non-vanishing torsion pseudo-vector $t_\lambda$. In particular, one can prove that the choice \eqref{choice} is the only one we can perform at this point in such a way to have non-trivial dynamics for $\phi$ in the presence of a non-vanishing (totally antisymmetric) torsion. Moreover, eq. \eqref{eqscal} reduces to
\begin{equation}
\frac{1}{\sqrt{-g}} \partial_\mu \left[ \sqrt{-g} \partial^\mu \phi \right] = \frac{\partial V}{\partial \phi} \,,
\end{equation}
which, in the simple case of a free scalar (i.e., $V(\phi)=0$), becomes
\begin{equation}\label{eqscalnew}
\frac{1}{\sqrt{-g}} \partial_\mu \left[ \sqrt{-g} \partial^\mu \phi \right] = 0 \,.
\end{equation}
Following the same lines of \cite{Iosifidis:2021kqo}, let us restrict ourselves to this case from now on. \\
Regarding the torsion, we are left with
\begin{equation}\label{torcomplas}
{S_{\lambda\mu}}^\nu = \frac{1}{6} \varepsilon_{\lambda \mu \kappa \rho} g^{\kappa \nu} t^\rho \,.
\end{equation}
Up to this point, the above considerations were general for the model at hand. Let us now focus on the homogeneous FLRW cosmology of this theory (for the cosmological setup we will consider in the following we refer the reader to Appendix \ref{appb}). From the first equation of \eqref{tornonmetFLRW} we see that, in this case, $\Phi(t)=0$. Hence, we find that the full torsion tensor is given by
\begin{equation}\label{formtor}
S_{\mu \nu \alpha} = \varepsilon_{\mu \nu \alpha \rho} u^\rho P(t) 
\end{equation}
in a homogeneous, non-Riemannian (metric but torsionful) FLRW cosmological setup.
Besides, plugging back the above results into the action \eqref{scal1}, the latter becomes
\begin{equation}\label{scal1new}
\mathcal{S} = \frac{1}{2 \kappa} \int \sqrt{-g} d^4 x \left[ R - T - \kappa g^{\mu \nu} \nabla_\mu \phi \nabla_\nu \phi \right] \,.
\end{equation}
Now, eq. \eqref{eqscalnew} implies
\begin{equation}\label{dotphieq}
\dot{\phi} = \frac{c_0}{a^3} \,,
\end{equation}
where $c_0$ is an arbitrary constant and $a=a(t)$ is the scale factor of the universe. \\
On the other hand, the metric field equations of the theory \eqref{scal1new} read
\begin{equation}\label{deltagscal1}
- \frac{1}{2} g_{\mu \nu} (R-T) + R_{(\mu \nu)} - \left( 2 S_{\nu \alpha \beta} {S_\mu}^{\alpha \beta} - S_{\alpha \beta \mu} {S^{\alpha \beta}}_\nu + 2 S_{\nu \alpha \beta} {S_\mu}^{\beta \alpha} \right) = \kappa T_{\mu \nu} \,.
\end{equation}
Here, let us introduce the following form of the energy momentum-tensor:
\begin{equation}\label{enmomtensscal}
T_{\mu \nu} = \rho u_\mu u_\nu + p h_{\mu \nu} \,,
\end{equation}
where $\rho$ and $p$ are, respectively, the density and the pressure associated with the scalar field Lagrangian, while $u^\mu$ is the normalized four-velocity and $h_{\mu \nu}$ is the projection tensor projecting objects on the space orthogonal to $u^\mu$ (cf. the definition in eq. \eqref{projop}).
Therefore, we have
\begin{equation}
\mathcal{T} = - \rho + 3 p = \dot{\phi}^2 \,.
\end{equation}
Taking the trace of \eqref{deltagscal1} and using \eqref{formtor} and \eqref{prexpcosm}, we get
\begin{equation}\label{fried1scal1}
\frac{\ddot{a}}{a} + \left(\frac{\dot{a}}{a} \right)^2 = - \frac{\kappa}{6} \dot{\phi}^2 \,.
\end{equation}
Let us now derive the acceleration equation for the case at hand. 
Contracting the field equations \eqref{deltagscal1} with $u^\mu u^\nu$, using the useful formulas collected in Appendix \ref{appb}, we obtain
\begin{equation}
R_{\mu \nu} u^\mu u^\nu = \frac{\kappa}{2} \left( \rho + 3 p \right) \,.   
\end{equation}
The latter, when substituted in eq. \eqref{secondF} in the case of vanishing non-metricity and $\Phi=0$, yields
\begin{equation}\label{acceleqscal0}
\frac{\ddot{a}}{a} = - \frac{\kappa}{6} \left( \rho + 3 p \right) \,,
\end{equation}
which is the acceleration equation of the theory. It can be rewritten as
\begin{equation}\label{acceleqscal}
\frac{\ddot{a}}{a} = - \frac{\kappa}{3} \dot{\phi}^2 \,. 
\end{equation}
Notice that the non-Riemannian degrees of freedom of the torsion pseudo-vector do not affect the cosmological evolution in this case, as all the contributions in $P(t)$ cancel out in the above equations.
Finally, observe that, using \eqref{acceleqscal} to eliminate the double temporal derivative of the scale factor in \eqref{fried1scal1}, the latter becomes
\begin{equation}\label{firstfriedfinal}
\left( \frac{\dot{a}}{a} \right)^2 = \frac{\kappa}{6} \dot{\phi}^2 \,,
\end{equation}
which is the first Friedmann equation of the model. Therefore, one is left with the usual scalar field cosmology, namely the one that would have been obtained by considering the purely Riemannian case.
The situation would obviously be different in the presence of non-vanishing scalar potential and non-metricity. In the present case, the torsion pseudo-vector, that here is the only non-vanishing part of the torsion tensor, do not affect the cosmological equations of the theory, and we end up with a well-known result. This is essentially a consequence of the fact that we have to fix the coupling constant $\lambda_0 = 0$ and the parameter $\beta=-1$ in order for the scalar field to have non-trivial dynamics and for the torsion to be non-vanishing, respectively. \\
The final form of the (metric but torsionful) affine connection for this model reads as follows:
\begin{equation}\label{connfincase1}
{\Gamma^\lambda}_{\mu \nu} = \tilde{\Gamma}^\lambda_{\phantom{\lambda}\mu \nu} + \frac{1}{6} {\varepsilon^{\lambda}}_{\mu \nu \rho} t^\rho \,,
\end{equation}
where $\tilde{\Gamma}^\lambda_{\phantom{\lambda}\mu \nu}$ is the Levi-Civita connection, defined in \eqref{lcconn}. Note that, using \eqref{formtor}, the final form of the connection given in \eqref{connfincase1} can be rewritten as
\begin{equation}
{\Gamma^\lambda}_{\mu \nu} = \tilde{\Gamma}^\lambda_{\phantom{\lambda}\mu \nu} + {\varepsilon^{\lambda}}_{\mu \nu \rho} u^\rho W(t) \,,
\end{equation}
with 
\begin{equation}
W(t)=P(t) \,.
\end{equation}
Let us conclude this section by observing that in the sub-case in which $F=R+\beta T \to F=R$ one would obtain a completely vanishing torsion tensor and the scalar field cosmology would be the same retrieved above. On the other hand, in the sub-case $F=R+\beta T \to F=T$, where the dependence on the scalar curvature $R$ in $F$ is removed and $\beta=1$ fixes the normalization of the theory, one is left with an imaginary coupling constant $\lambda_0 = \pm 2 \ii \sqrt{\frac{2}{3 \kappa}}$. For the sake of completeness, we report this particular sub-case in Appendix \ref{FTplusscal}.

\section{Conclusions}\label{concl}

This review, where we have collected the MG-N models and their MAMG generalizations, including also their cosmological analysis, will be instrumental to the reader interested in these topics and also to further investigate mathematical, physical, and cosmological aspects and applications of the aforementioned theories, in view of possible comparisons with observations.\footnote{For some result that has already been obtained in this context we refer the reader to \cite{Anagnostopoulos:2020lec}.} \\
We have discussed the MAMG-N theories, offering a complete dictionary on the subject, and derived the modified Friedmann equations for the linear case in a homogeneous, non-Riemannian FLRW spacetime. In this context, let us stress that it would be interesting to release the assumptions of linearity of $F$, in particular to examine the cosmological consequences induced by the very presence of the divergence of the dilation current in $F$. Such a study might also shed some light on the phenomenological aspects of this peculiar contribution and on the energy-momentum trace counterpart as well. \\
Besides, we have given a cosmological application of the results obtained so far to the case in which the matter action is given in terms of the Lagrangian for a scalar field directly coupled to torsion by means of the torsion vector $S^\mu$, with coupling constant $\lambda_0$. In particular, focusing on the linear metric-affine $F=R+\beta T$ theory (i.e., the linear MAMG-I model), we have shown that we actually have to set $\lambda_0=0$ (and $\beta=-1$) in order to end up with a non-trivial theory. This implies that, in this case, the torsion does not contribute to the cosmological evolution of the model and one ends up with the usual scalar field cosmology. Nevertheless, the torsion is non-vanishing in this case: it is completely antisymmetric (given in terms of the torsion pseudo-vector) and determines the final form of the affine connection. \\
As already stressed in \cite{Iosifidis:2021kqo}, one could extend our analysis by considering also non-vanishing non-metricity and hence direct coupling of the scalar field to both torsion and non-metricity. In this context, one might take the following matter Lagrangian:
\begin{equation}\label{scalLagrnm}
\mathcal{L}_{\text{m}} = - \frac{1}{2} g^{\mu \nu} \nabla_\mu \phi \nabla_\nu \phi - V (\phi) + \lambda_0 S^\mu \nabla_\mu \phi + \lambda_1 Q^\mu \nabla_\mu \phi + \lambda_2 q^\mu \nabla_\mu \phi \,,
\end{equation}
with constant parameters $\lambda_0$, $\lambda_1$, $\lambda_2$ and where $Q^\mu$ and $q^\mu$ are the non-metricity vectors. In particular, the hypermomentum associated with \eqref{scalLagrnm} is\footnote{It can be derived by exploiting eqs. \eqref{variationQSpart}.}
\begin{equation}
{\Delta_\lambda}^{\mu \nu} = 2 \left[ \lambda_0 \delta_\lambda^{[\mu} \nabla^{\nu]} \phi - 2 \lambda_1 \delta_\lambda^\mu \nabla^\nu \phi - \lambda_2 \left( g^{\mu \nu} \nabla_\lambda \phi + \delta^\nu_\lambda \nabla^\mu \phi \right) \right] \,.
\end{equation}
In this case, we expect the non-metricity vectors, and possibly the torsion trace, to play a non-trivial role in cosmological solutions. We leave the complete study to a forthcoming paper.

It would be worth it to further elaborate on cosmological applications of MAMG theories, especially in the presence of a cosmological hyperfluid \cite{Iosifidis:2020gth,Iosifidis:2021nra} (see also \cite{Iosifidis:2021iuw}). Possible future developments may also regard the inclusion of parity violating terms in MAMG, in particular of the so-called Hojman term \cite{Hojman:1980kv} (most of the time referred to as the Holst term \cite{Holst:1995pc}), following the lines of \cite{Iosifidis:2020dck}. Always in this context, one might also consider the inclusion of a parity violating coupling of a scalar field to torsion, adding the contribution $\lambda_3 t^\mu \nabla_\mu \phi$ to the matter Lagrangian \eqref{scalLagrnm}, where $\lambda_3$ is another constant parameter of the theory. In this case, the hypermomentum tensor would be
\begin{equation}
{\Delta_\lambda}^{\mu \nu} = 2 \left[ \lambda_0 \delta_\lambda^{[\mu} \nabla^{\nu]} \phi - 2 \lambda_1 \delta_\lambda^\mu \nabla^\nu \phi - \lambda_2 \left( g^{\mu \nu} \nabla_\lambda \phi + \delta^\nu_\lambda \nabla^\mu \phi \right) - \lambda_3 {\varepsilon^{\rho \mu \nu}}_\lambda \nabla_\rho \phi \right] \,.
\end{equation}
Given this enlarged setup, we argue that here also the torsion pseudo-vector could play a non-trivial role in the study of cosmological solutions.

\section*{Author contributions}

Writing---original draft preparation, L.R.; writing---review and editing, N.M. and R.M.

\section*{Funding information}

This research received no external funding.

\section*{Conflict of interest}

The authors declare no conflict of interest.

\section*{Acknowledgments}

The work was supported by the Ministry of Education and Science of the Republic of Kazakhstan, Grant
AP09058240.
L.R. acknowledges the Department of Applied Science and Technology of the Polytechnic of Turin, and in particular Laura Andrianopoli and Francesco Raffa, for financial support.
We would like to thank Damianos Iosifidis for useful discussions.

\appendix

\section{Notation, conventions and useful formulas in MAG}\label{appa}

We consider four spacetime dimensions and our convention for the metric signature is mostly plus: $\left(-,+,+,+\right)$. We use minuscule Greek letters to denote spacetime indices, that is $\mu, \nu,\ldots=0,1,2,3$. \\
The covariant derivative $\nabla$ of, e.g., a vector $v^\lambda$ is defined as
\begin{equation}
\nabla_\nu v^\lambda = \partial_\nu v^\lambda  + {\Gamma^\lambda}_{\mu \nu} v^\mu \,,
\end{equation}
where we have also introduced the general affine connection ${\Gamma^\lambda}_{\mu \nu}$. The latter can be decomposed as follows:
\begin{equation}\label{gendecompaffconn}
{\Gamma^\lambda}_{\mu\nu} = \tilde{\Gamma}^\lambda_{\phantom{\lambda}\mu\nu} + {N^\lambda}_{\mu\nu}\,,
\end{equation}
where
\begin{equation}\label{distortion}
{N^\lambda}_{\mu\nu} = \underbrace{\frac12 g^{\rho\lambda}\left(Q_{\mu\nu\rho} + Q_{\nu\rho\mu}
- Q_{\rho\mu\nu}\right)}_{\text{deflection {(or disformation)}}} - \underbrace{g^{\rho\lambda}\left(S_{\rho\mu\nu} +
S_{\rho\nu\mu} - S_{\mu\nu\rho}\right)}_{\text{contorsion} \, := \, {K^\lambda}_{\mu \nu}}
\end{equation}
is the distortion tensor and
\begin{equation}\label{lcconn}
\tilde{\Gamma}^\lambda_{\phantom{\lambda}\mu\nu} = \frac12 g^{\rho\lambda}\left(\partial_\mu 
g_{\nu\rho} + \partial_\nu g_{\rho\mu} - \partial_\rho g_{\mu\nu}\right)
\end{equation}
is the Levi-Civita connection. The torsion and non-metricity tensors in \eqref{distortion} are defined as
\begin{equation}
\begin{split}
{S_{\mu\nu}}^\rho & := {\Gamma^\rho}_{[\mu\nu]}\,, \\ 
Q_{\lambda\mu\nu} & := -\nabla_\lambda g_{\mu\nu} = 
-\partial_\lambda g_{\mu\nu} + {\Gamma^\rho}_{\mu\lambda} g_{\rho\nu} +
{\Gamma^\rho}_{\nu\lambda}g_{\mu\rho} \,,
\end{split}
\end{equation}
respectively. Their trace decomposition in four spacetime dimensions reads
\begin{equation}\label{dectorandnm}
\begin{split}
{S_{\lambda\mu}}^\nu & = \frac{2}{3} {\delta_{[\mu}}^{\nu} S_{\lambda]} + \frac{1}{6} \varepsilon_{\lambda \mu \kappa \rho} g^{\kappa \nu} t^\rho + {Z_{\lambda\mu}}^\nu \,,  \\
Q_{\lambda\mu\nu} & = \frac{5}{18} Q_\lambda g_{\mu\nu} - \frac19 q_\lambda g_{\mu\nu} +
\frac49 g_{\lambda(\nu}q_{\mu)} - \frac19 g_{\lambda(\nu} Q_{\mu)} + \Omega_{\lambda\mu\nu} \,, 
\end{split}
\end{equation}
where $Q_\lambda := {Q_{\lambda \mu}}^\mu$ and $q_\nu := {Q^\mu}_{\mu\nu}$ are the non-metricity vectors, $S_\lambda :={S_{\lambda \sigma}}^{\sigma}$ is the torsion vector, and $t^\rho := \varepsilon^{\rho \lambda \mu \nu} S_{\lambda \mu \nu}$ is the torsion pseudo-vector. On the other hand, ${Z_{\lambda\mu}}^\nu$ (with $Z_{\lambda \mu \nu} = \frac{4}{3} Z_{[\lambda (\mu]\nu)}$, $\epsilon^{\lambda \mu \nu \rho} Z_{\lambda \mu \nu}=0$) and $\Omega_{\lambda\mu\nu}$ are the traceless parts of torsion and non-metricity, respectively.
We denote by $\varepsilon^{\mu \nu \alpha \beta}= \frac{1}{\sqrt{-g}} \epsilon^{\mu \nu \alpha \beta}$ the Levi-Civita tensor, while $\epsilon^{\mu \nu \alpha \beta}$ is the Levi-Civita symbol. \\
Exploiting the trace decomposition \eqref{dectorandnm}, one can prove that the so-called Palatini tensor, whose definition reads
\begin{equation}\label{palatinidefin}
{P_{\lambda}}^{\mu\nu} := -\frac{\nabla_{\lambda}(\sqrt{-g}g^{\mu\nu})}{\sqrt{-g}}+\frac{\nabla_{\sigma}(\sqrt{-g}g^{\mu\sigma})\delta^{\nu}_{\lambda}}{\sqrt{-g}} +2(S_{\lambda}g^{\mu\nu}-S^{\mu}\delta_{\lambda}^{\nu}+g^{\mu\sigma}{S_{\sigma\lambda}}^{\nu}) \,,
\end{equation}
can be written in terms of torsion and non-metricity as \cite{Iosifidis:2019jgi}
\begin{equation}\label{palatinitornonmet}
\begin{split}
{P_{\lambda}}^{\mu\nu} & = \delta^\nu_\lambda \left(q^\mu - \frac{1}{2} Q^\mu - 2 S^\mu \right) + g^{\mu \nu} \left( \frac{1}{2} Q_\lambda + 2 S_\lambda \right) - \left( {Q_\lambda}^{\mu \nu} + 2 {S_\lambda}^{\mu \nu} \right) \\
& = - {\Omega_\lambda}^{\mu \nu} + \frac{1}{3} g^{\mu \nu} \left( \frac{2}{3} Q_\lambda + \frac{1}{3} q_\lambda + 4 S_\lambda \right) + \frac{1}{9} \delta_\lambda^\nu \left(-4 Q^\mu + 7 q^\mu \right) + \frac{1}{9} \delta_\lambda^\mu  \left( \frac{1}{2} Q^\mu - 2 q^\nu \right) \\
& - \frac{1}{3} {\varepsilon_\lambda}^{\mu \nu \rho} t_\rho - 2 {Z_\lambda}^{\mu \nu} \,.
\end{split}
\end{equation}
Note that the Palatini tensor is traceless in the indices $\mu,\lambda$, that is
\begin{equation}\label{palatinitraceless}
{P_{\mu}}^{\mu\nu}=0 \,.
\end{equation}
Our definition of the Riemann tensor for the general affine connection ${\Gamma^\lambda}_{\mu \nu}$ is
\begin{equation}\label{defRiem}
{R^\mu}_{\nu \alpha \beta} := 2 \partial_{[\alpha} {\Gamma^\mu}_{|\nu|\beta]} + 2 {\Gamma^\mu}_{\rho [\alpha} {\Gamma^\rho}_{|\nu |\beta]} \,.
\end{equation}
Correspondingly, $R_{\mu \nu}:={R^\rho}_{\mu \rho \nu}$ and $R:=g^{\mu \nu} R_{\mu \nu}$ are, respectively, the Ricci tensor and the scalar curvature of $\Gamma$.
In four spacetime dimensions the Riemann tensor in \eqref{defRiem} can be decomposed into its Riemannian and non-Riemannian parts as follows:
\begin{equation}\label{RiemdecLCTNM}
\begin{split}
R_{\lambda \mu \nu \kappa} & = \frac{1}{2} \left(  g_{\lambda \nu} \tilde{R}_{\mu \kappa} - g_{\lambda \kappa} \tilde{R}_{\mu \nu} - g_{\mu \nu} \tilde{R}_{\lambda \kappa} + g_{\mu \kappa} \tilde{R}_{\lambda \nu} \right) - \frac{1}{6} \tilde{R} \left( g_{\lambda \nu} g_{\mu \kappa} - g_{\lambda \kappa} g_{\mu \nu} \right) + C_{\lambda \mu \nu \kappa} \\
& + \tilde{\nabla}_\kappa N_{\lambda \mu \nu} - \tilde{\nabla}_\nu N_{\lambda \mu \kappa} + N_{\lambda \alpha \nu} {N^\alpha}_{\mu \kappa} - N_{\lambda \alpha \kappa} {N^\alpha}_{\mu \nu} \,,
\end{split}
\end{equation}
where $\tilde{R}_{\mu \nu}$ and $\tilde{R}:=g^{\mu \nu} \tilde{R}_{\mu \nu}$ are, respectively, the Ricci tensor and Ricci scalar of the Levi-Civita connection $\tilde{\Gamma}^\lambda_{\phantom{\lambda}\mu \nu}$, $\tilde{\nabla}$ denotes the covariant derivative of $\tilde{\Gamma}^\lambda_{\phantom{\lambda}\mu \nu}$, and ${C^\lambda}_{\mu \nu \kappa}$ is the Weyl tensor, fulfilling
\begin{equation}
{C^\lambda}_{\mu \lambda \kappa} = 0 \,, \quad C^{(\lambda \mu) \nu \rho} = 0 \,, \quad C^{\lambda \mu (\nu \rho)} = 0 \,, \quad C^{\lambda \mu \nu \rho} = C^{\nu \rho \lambda \mu} \,, \quad C^{[\lambda \mu \nu \rho]} = 0 \,, \quad g_{\rho \nu} C^{\rho \lambda \mu \nu} = 0 \,.
\end{equation}
Furthermore, any tensor contraction between indices of the Weyl tensor vanishes. \\
We can also introduce the decomposition of the scalar curvature $R$ in terms of the Riemannian scalar curvature $\tilde{R}$ plus the non-Riemannian contributions given by torsion and non-metricity scalars, that is
\begin{equation}\label{ricciscaldec}
R = \tilde{R} + T + Q + 2 Q_{\alpha \mu \nu} S^{\alpha \mu \nu} + 2 S_\mu \left( q^\mu - Q^\mu \right) + \tilde{\nabla} \left( q^\mu - Q^\mu - 4 S^\mu \right) \,,
\end{equation}
where we have defined the torsion and non-metricity scalars as
\begin{equation}\label{tornonmetscals}
\begin{split}
T & := S_{\mu \nu \alpha} S^{\mu \nu \alpha} - 2 S_{\mu \nu \alpha} S^{\alpha \mu \nu} - 4 S_\mu S^\mu \,, \\
Q & := \frac{1}{4} Q_{\alpha \mu \nu} Q^{\alpha \mu \nu} - \frac{1}{2} Q_{\alpha \mu \nu} Q^{\mu \nu \alpha} - \frac{1}{4} Q_\mu Q^\mu + \frac{1}{2} Q_\mu q^\mu \,,
\end{split}
\end{equation}
respectively. Moreover, notice that defining the ``superpotentials''
\begin{equation}\label{superpot}
\begin{split}
\Xi^{\alpha \mu \nu} & := \frac{1}{4} Q^{\alpha \mu \nu} - \frac{1}{2} Q^{\mu \nu \alpha} - \frac{1}{4} g^{\mu \nu} Q^\alpha + \frac{1}{2} g^{\alpha \mu} Q^\nu \,, \\
\Sigma^{\alpha \mu \nu} & := S^{\alpha \mu \nu} - 2 S^{\mu \nu \alpha} - 4 g^{\mu \nu} S^\alpha 
\end{split}
\end{equation}
the torsion and non-metricity scalars in \eqref{tornonmetscals} can be rewritten in the following, more compact, form:
\begin{equation}
\begin{split}
T & = S_{\alpha \mu \nu} \Sigma^{\alpha \mu \nu} \,, \\
Q & = Q_{\alpha \mu \nu} \Xi^{\alpha \mu \nu} \,.
\end{split}
\end{equation}
Let us also report the useful formulas concerning the variation of the non-metricity and torsion tensors with respect to the metric and the general affine connection ${\Gamma^\lambda}_{\mu \nu}$, that is \cite{Iosifidis:2019jgi}
\begin{equation}\label{variationQS}
\begin{split}
& \delta_g Q_{\rho\alpha\beta} = \partial_\rho\left(g_{\mu\alpha} g_{\nu\beta} \delta g^{\mu\nu}\right) - 
2 g_{\lambda\mu} g_{\nu(\alpha} {\Gamma^\lambda}_{\beta)\rho}\delta g^{\mu\nu}\,, \\
& \delta_g {S_{\mu\nu}}^\alpha = 0 \,, \\
& \delta_\Gamma Q_{\rho\alpha\beta} = 2\delta^\nu_\rho\delta^\mu_{(\alpha} g_{\beta)\lambda}
\delta {\Gamma^\lambda}_{\mu\nu}\,, \\
& \delta_\Gamma {S_{\alpha\beta}}^\lambda = \delta^{[\mu}_{\alpha}\delta^{\nu]}_{\beta}
\delta {\Gamma^\lambda}_{\mu\nu}\,.
\end{split}
\end{equation}
In particular, it follows that
\begin{equation}\label{variationQSpart}
\begin{split}
& \delta_g Q_\rho = \partial_\rho \left( g_{\mu \nu} \delta g^{\mu \nu} \right) \,, \\
& \delta_g q_\beta = \delta g^{\mu \nu} \left[ g_{\nu \beta} g^{\rho \alpha} \left( \partial_\rho g_{\mu \alpha} \right) + {\Gamma^\lambda}_{\mu \nu} g_{\lambda \beta} - g^{\rho \sigma} {\Gamma^\alpha}_{\rho \sigma} g_{\mu \alpha} g_{\nu \beta} \right] + g_{\nu \beta} \left(\partial_\mu \delta g^{\mu \nu} \right) \,, \\
& \delta_g S_\mu = 0 \,, \\
& \delta_\Gamma Q_\rho = 2 \delta^\nu_\rho \delta^\mu_\lambda \delta {\Gamma^\lambda}_{\mu \nu} \,, \\
& \delta_\Gamma q_\beta = \left( g^{\mu \nu} g_{\beta \lambda} + \delta^\mu_\beta \delta^\nu_\lambda \right) \delta {\Gamma^\lambda}_{\mu \nu} \,, \\
& \delta_\Gamma S_\alpha = \delta^{[\mu}_\alpha \delta^{\nu]}_\lambda \delta {\Gamma^\lambda}_{\mu \nu} \,, \\
& \delta_\Gamma t^\alpha = {\varepsilon^{\alpha \mu \nu}}_\lambda \delta {\Gamma^\lambda}_{\mu \nu} \,.
\end{split}
\end{equation}
Finally, regarding the matter content in MAG theories, besides the usual energy-momentum
tensor, which is defined as
\begin{equation}
T_{\mu \nu} := - \frac{2}{\sqrt{-g}} \frac{\delta \left( \sqrt{-g} \mathcal{L}_{\text{m}} \right)}{\delta g^{\mu \nu}} \,,
\end{equation}
we also have a non-trivial dependence of the matter Lagrangian on the general affine connection. In fact, the variation of the matter part of the action with respect to ${\Gamma^\lambda}_{\mu \nu}$ defines the hypermomentum tensor,
\begin{equation}
{\Delta_\lambda}^{\mu \nu} := - \frac{2}{\sqrt{-g}} \frac{\delta \left( \sqrt{-g} \mathcal{L}_{\text{m}}\right)}{\delta {\Gamma^\lambda}_{\mu \nu}} \,.
\end{equation}
The energy-momentum and hypermomentum tensors are not independent. In particular, they are subject to the conservation law
\begin{equation}\label{conslawenmomhyperm}
\sqrt{-g} \left( 2 \tilde{\nabla}_\mu {T^\mu}_\alpha - \Delta^{\lambda \mu \nu} R_{\lambda \mu \nu \alpha} \right) + \check{\nabla}_\mu \check{\nabla}_\nu \left( \sqrt{-g} {\Delta_\alpha}^{\mu \nu} \right) + 2 {S_{\mu \alpha}}^\lambda \check{\nabla}_\nu \left( \sqrt{-g} {\Delta_\lambda}^{\mu \nu} \right) = 0 \,, 
\end{equation}
where we have introduced
\begin{equation}
\check{\nabla}_\mu := 2 S_\mu - \nabla_\mu = \sqrt{-g} \hat{\nabla} \,,
\end{equation}
the derivative $\hat{\nabla}$ being defined in \eqref{hatder}.
Eq. \eqref{conslawenmomhyperm} originates from the invariance under diffeomorphisms of the matter sector of the action (cf. \cite{Iosifidis:2020gth}).

\section{Cosmology in the presence of torsion and non-metricity}\label{appb}

In this appendix we recall, following \cite{Iosifidis:2020gth}, the expressions of the general affine connection ${\Gamma^\lambda}_{\mu \nu}$, the torsion and non-metricity tensors, the Palatini tensor, and the scalar curvature $R$ in a homogeneous, non-Riemannian FLRW spacetime. \\
We consider a homogeneous, flat FLRW cosmology, where the line element is
\begin{equation}
ds^2 = - dt^2 + a^2(t) \delta_{ij} dx^i dx^j \,,
\end{equation}
with $i,j,\ldots=1,2,3$ and scale factor $a(t)$. The Hubble parameter is
\begin{equation}
H:= \frac{\dot{a}}{a}
\end{equation}
and the projection tensor projecting objects on the space orthogonal to the normalized four-velocity $u^\mu$ (such that $u^\mu=\delta^\mu_0=(1,0,0,0)$ and $u_\mu u^\mu=-1$) is
\begin{equation}\label{projop}
h_{\mu \nu} := g_{\mu \nu} + u_\mu u_\nu = h_{\nu \mu} \,.
\end{equation}
In particular, we have
\begin{equation}
h^{\mu \nu} h_{\mu \nu} = 3 \,, \quad h_{\mu \alpha} h^{\nu \alpha} = \delta^\nu_\mu + u_\mu u^\nu \,, \quad {h^\mu}_\mu = 3 \,, \quad h_{\mu \nu}u^\mu u^\nu = 0 \,.
\end{equation}
We also define the temporal derivative
\begin{equation}\label{tempder}
\dot{} = u^\alpha \nabla_\alpha \,.
\end{equation}
The projection operator \eqref{projop} and the temporal derivative \eqref{tempder} constitute together a $1 + 3$ spacetime split. \\
In a non-Riemannian FLRW spacetime in $1+3$ dimensions the general affine connection can be written as 
\begin{equation}\label{connFLRW}
{\Gamma^\lambda}_{\mu \nu} = \tilde{\Gamma}^\lambda_{\phantom{\lambda}\mu \nu} + X(t) u^\lambda h_{\mu \nu} + Y(t) u_\mu {h^\lambda}_\nu + Z(t) u_\nu {h^\lambda}_\mu + V(t) u^\lambda u_\mu u_\nu + {\varepsilon^\lambda}_{\mu \nu \rho} u^\rho W(t) \,,
\end{equation}
where, in particular, the non-vanishing components of the Levi-Civita connection read
\begin{equation}
\tilde{\Gamma}^0_{\phantom{0}ij} = \tilde{\Gamma}^0_{\phantom{0}ji} = \dot{a} a \delta_{ij} = H g_{ij} \,, \quad \tilde{\Gamma}^i_{\phantom{i}j0} = \tilde{\Gamma}^i_{\phantom{i}0j} = \frac{\dot{a}}{a} \delta^i_j = H \delta^i_j \,. 
\end{equation}
The torsion and non-metricity tensors can be written, respectively, in the following way:
\begin{equation}\label{tornonmetFLRW}
\begin{split}
S_{\mu \nu \alpha} & = 2 u_{[\mu} h_{\nu]\alpha} \Phi(t) + \varepsilon_{\mu \nu \alpha \rho} u^\rho P(t) \,, \\
Q_{\alpha \mu \nu} & = A(t) u_\alpha h_{\mu \nu} + B(t) h_{\alpha(\mu} u_{\nu)} + C(t) u_\alpha u_\mu u_\nu \,.
\end{split}
\end{equation}
The functions $X(t)$, $Y(t)$, $Z(t)$, $V(t)$, $W(t)$ in \eqref{connFLRW} and $\Phi(t)$, $P(t)$, $A(t)$, $B(t)$, $C(t)$ in \eqref{tornonmetFLRW} describe non-Riemannian cosmological effects and give, together with the scale factor, the cosmic evolution of non-Riemannian geometries. 
Moreover, recalling \eqref{gendecompaffconn} and plugging \eqref{connFLRW} and \eqref{tornonmetFLRW} into \eqref{distortion}, one can prove that the following linear relations hold among the functions introduced above:
\begin{equation}
2(X+Y) = B \,, \quad 2 Z = A \,, \quad 2 V = C \,, \quad 2 \Phi = Y - Z \,, \quad P=W \,,
\end{equation}
which may also be inverted, obtaining
\begin{equation}
W = P \,, \quad V= \frac{C}{2} \,, \quad Z = \frac{A}{2} \,, \quad Y = 2 \Phi + \frac{A}{2} \,, \quad X = \frac{B}{2} - 2 \Phi - \frac{A}{2} \,.
\end{equation}
Therefore, one can prove that the torsion and non-metricity scalars defined in \eqref{tornonmetscals} become, respectively,
\begin{equation}\label{tornmscalFLRM}
\begin{split}
T & = 24 \Phi^2 - 6 P^2 \,, \\
Q & = \frac{3}{4} \left[ 2 A^2 + B (C-A) \right] \,.
\end{split}
\end{equation}
Besides, let us also mention that, using \eqref{tornonmetFLRW} into the explicit expression of the Palatini tensor in terms of torsion and non-metricity, that is \eqref{palatinitornonmet}, we get the cosmological expression of the Palatini tensor, which reads
\begin{equation}
\begin{split}
P_{\alpha \mu \nu} & = \left( \frac{1}{2} A + 4 \Phi - \frac{C}{2} \right) u_\alpha h_{\mu \nu} + \left( B - \frac{3}{2} A - 4 \Phi - \frac{C}{2} \right) u_\mu h_{\alpha \nu} - \frac{B}{2} u_\nu h_{\mu \alpha} - \frac{3}{2} B u_\alpha u_\mu u_\nu \\
& - 2 \varepsilon_{\alpha \mu \nu \rho} u^\rho P \,,
\end{split}
\end{equation}
or, in the form we use in the main text of this paper,
\begin{equation}\label{cosmPala}
\begin{split}
{P_{\lambda}}^{\mu \nu} & = \left( \frac{1}{2} A + 4 \Phi - \frac{C}{2} \right) u_\lambda h^{\mu \nu} + \left( B - \frac{3}{2} A - 4 \Phi - \frac{C}{2} \right) u^\mu {h_\lambda}^\nu - \frac{B}{2} u^\nu {h^\mu}_\lambda - \frac{3}{2} B u_\lambda u^\mu u^\nu \\
& - 2 \varepsilon_{\lambda \phantom{\mu \nu} \rho}^{\phantom{\lambda} \mu \nu} u^\rho P \,.
\end{split}
\end{equation}
Consequently, one can prove that the following relations hold:
\begin{equation}\label{cosmPalacontr}
\begin{split}
& h^{\alpha \mu} P_{\alpha \mu \nu} = - \frac{3}{2} B u_\nu \,, \\
& h^{\alpha \nu}P_{\alpha \mu \nu} = 3 \left( B - \frac{3}{2} A - 4 \Phi - \frac{C}{2} \right) u_\mu \,, \\
& h^{\mu \nu} P_{\alpha \mu \nu} = 3 \left( \frac{1}{2} A + 4 \Phi - \frac{C}{2} \right) u_\alpha \,, \\
& \varepsilon^{\alpha \mu \nu \lambda} P_{\alpha \mu \nu} = 12 P u^\lambda \,, \\
& u^\alpha u^\mu u^\nu P_{\alpha \mu \nu} = \frac{3}{2} B \,.
\end{split}
\end{equation}
Finally, using \eqref{tornmscalFLRM}, we find that the scalar curvature $R$, once decomposed in its Riemannian and non-Riemannian parts (see eq. \eqref{ricciscaldec}), acquires the following form:
\begin{equation}\label{prexpcosm}
R = \tilde{R} + 6 \left[ \frac{1}{4} A^2 + 4 \Phi^2 + \Phi (2A-B) \right] + \frac{3}{4} B (C-A) - 6 P^2 + \frac{3}{\sqrt{-g}} \partial_\mu \left[ \sqrt{-g} u^\mu \left( \frac{B}{2} - A - 4 \Phi \right) \right] \,,
\end{equation}
where
\begin{equation}
\tilde{R} = 6 \left[ \frac{\ddot{a}}{a} + \left( \frac{\dot{a}}{a} \right)^2 \right]
\end{equation}
is the Ricci scalar of the Levi-Civita connection $\tilde{\Gamma}^\lambda_{\phantom{\lambda} \mu \nu}$.

\section{Cosmological aspects of the metric-affine $F(T)$ theory with a scalar field coupled to torsion}\label{FTplusscal}

In this appendix we take the model of Subsection \ref{applic} and restrict ourselves to the sub-case $F=R+\beta T \to F=T$ in the presence of a (free) scalar field coupled to torsion, namely we consider the action
\begin{equation}\label{scal2}
\mathcal{S} = \frac{1}{2 \kappa} \int \sqrt{-g} d^4 x \left[ T + 2 \kappa \left( - \frac{1}{2} g^{\mu \nu} \nabla_\mu \phi \nabla_\nu \phi + \lambda_0 S^\mu \nabla_\mu \phi \right) \right] \,.
\end{equation}
The variation of \eqref{scal2} with respect to the scalar field $\phi$ yields
\begin{equation}\label{eqscal2}
\frac{1}{\sqrt{-g}} \partial_\mu \left[ \sqrt{-g} \left( \partial^\mu \phi - \lambda_0 S^\mu \right) \right] = 0 \,.
\end{equation}
The hypermomentum tensor is given by \eqref{hypermscal} and, varying the action \eqref{scal2} with respect to the general affine connection ${\Gamma^\lambda}_{\mu \nu}$, we obtain
\begin{equation}\label{connfescal2}
2 \left({S^{\mu \nu}}_\lambda - 2 {S_\lambda}^{[\mu \nu]} - 4 S^{[\mu} \delta^{\nu]}_\lambda \right) = 2 \kappa \lambda_0 \delta^{[\mu}_\lambda \nabla^{\nu]} \phi \,.
\end{equation}
Taking the different contractions of \eqref{connfescal2}, one can prove that the latter is solved by
\begin{equation}\label{correctres2}
\begin{split}
& S^\mu = \frac{3 \kappa \lambda_0}{8} \partial^\mu \phi \,, \\
& t_\lambda = 0 \,, \quad {Z_\lambda}^{\mu \nu} = 0 \,.
\end{split}
\end{equation}
Hence, only the torsion vector survives and it is a pure gauge, while the torsion pseudo-vector and the traceless part of the torsion vanish. In particular, the first of \eqref{correctres2} indicates that it is the presence of the scalar field that produces spacetime torsion in this model. This is so because we have a non-vanishing hypermomentum. \\
Plugging this result back into the action \eqref{scal2}, we end up with
\begin{equation}\label{scal2new}
\mathcal{S} = \frac{1}{2 \kappa} \int \sqrt{-g} d^4 x \left[ T - \kappa \left( 1 - \frac{3 \kappa \lambda_0^2}{4} \right) g^{\mu \nu} \nabla_\mu \phi \nabla_\nu \phi \right] \,.
\end{equation}
Interestingly, from eq. \eqref{scal2new} we can conclude that the interaction between the scalar and torsion modifies the factor of the kinetic term for the scalar field. Observe also that there is a peculiar value of the coupling constant, namely $|\lambda_0|=2 \sqrt{\frac{1}{2 \kappa}}$, for which the kinetic
term of the scalar disappears from \eqref{scal2new}. Of course, we shall disregard this trivial case in the following. \\
Up to this point, the above considerations were general for the model at hand. We shall now focus on the homogeneous FLRW cosmology of this theory (c.f. Appendix \ref{appb} for the setup, definitions and useful formulas). Comparing the first equation of \eqref{correctres2} with the first of \eqref{tornonmetFLRW}, we immediately see that, in this case, $P(t)=0$. Hence, we find that the full torsion tensor is given by
\begin{equation}\label{formtornew}
S_{\mu \nu \alpha} = 2 u_{[\mu} h_{\nu]\alpha} \Phi \,,
\end{equation}
with
\begin{equation}\label{Phiexpr}
\Phi = \Phi(t) = - \frac{\kappa \lambda_0}{8} \dot{\phi} \,.
\end{equation}
Besides, inserting the first of \eqref{correctres2} into eq. \eqref{eqscal2}, we obtain
\begin{equation}\label{eqscal2new}
\left( 1 - \frac{3 \kappa \lambda_0^2}{4} \right) \partial_\mu \left[ \sqrt{-g} \partial^\mu \phi \right] = 0 \,,
\end{equation}
which, for $|\lambda_0| \neq 2 \sqrt{\frac{1}{2 \kappa}} $, implies \eqref{dotphieq}. \\
On the other hand, the metric field equations of the theory \eqref{scal1new} read
\begin{equation}\label{deltagscal2}
- \frac{1}{2} g_{\mu \nu} T + \left( 2 S_{\nu \alpha \beta} {S_\mu}^{\alpha \beta} - S_{\alpha \beta \mu} {S^{\alpha \beta}}_\nu + 2 S_{\nu \alpha \beta} {S_\mu}^{\beta \alpha} \right) = \kappa T_{\mu \nu} \,.
\end{equation}
Introducing the usual form of the energy-momentum tensor, that is \eqref{enmomtensscal}, associated with the scalar field Lagrangian, taking the trace of \eqref{deltagscal2}, and using eq. \eqref{formtornew}, we get (recall that we are not dealing with a gauge theory of gravity)
\begin{equation}\label{fried1scal2}
\left[ \left(\frac{\kappa \lambda_0}{4} \right)^2 + \frac{\kappa}{6} \right] \dot{\phi}^2 = 0 \,,
\end{equation}
which, discarding the (trivial) case $\dot{\phi}^2=0$, is solved by
\begin{equation}\label{l0form}
\lambda_0 = \pm 2 \ii \sqrt{\frac{2}{3 \kappa}} \,.
\end{equation}
Let us also observe that, contracting eq. \eqref{deltagscal2} with $u^\mu u^\nu$, we find the same equation \eqref{fried1scal2}. Thus, in the current model one is left with an imaginary coupling constant $\lambda_0$. \\
The final form of the (metric but torsionful) affine connection for the model at hand reads as follows:
\begin{equation}\label{connfincase2}
{\Gamma^\lambda}_{\mu \nu} = \tilde{\Gamma}^\lambda_{\phantom{\lambda}\mu \nu} - \frac{2}{3} g_{\mu \nu} S^\lambda + \frac{2}{3} \delta_\nu^\lambda S_\mu \,,
\end{equation}
where $\tilde{\Gamma}^\lambda_{\phantom{\lambda}\mu \nu}$ is the Levi-Civita connection. Note that, using \eqref{formtornew}, the final form of the connection in \eqref{connfincase2} can be rewritten as
\begin{equation}
{\Gamma^\lambda}_{\mu \nu} = \tilde{\Gamma}^\lambda_{\phantom{\lambda}\mu \nu} + X(t) u^\lambda h_{\mu \nu} + Y(t) u_\mu {h^\lambda}_\nu \,,
\end{equation}
with 
\begin{equation}
X(t) = - Y(t) = - 2 \Phi(t) \,.
\end{equation}
Finally, exploiting \eqref{Phiexpr}, we are left with
\begin{equation}
{\Gamma^\lambda}_{\mu \nu} = \tilde{\Gamma}^\lambda_{\phantom{\lambda}\mu \nu} + \frac{\kappa \lambda_0}{4} \dot{\phi} \left( u^\lambda h_{\mu \nu} - u_\mu {h^\lambda}_\nu \right) \,,
\end{equation}
written in terms of the temporal derivative of the scalar, $\dot{\phi}$. In fact, in this case, it is the scalar field that produces spactime torsion (cf. the first equation in \eqref{correctres2}).

\end{document}